\documentclass[acmsmall]{acmart}

\setcopyright{acmlicensed}
\acmJournal{PACMHCI}
\acmYear{2025} \acmVolume{9} \acmNumber{GROUP}
\acmArticle{009} \acmMonth{1} \acmPrice{15.00}
\acmDOI{10.1145/3701188}

\usepackage{xcolor}
\usepackage{tcolorbox}

\definecolor{pastelgreen}{RGB}{193, 225, 193}
\definecolor{pastelblue}{RGB}{192, 192, 255}
\definecolor{pastelpurple}{RGB}{220, 192, 255}
\definecolor{pastelyellow}{RGB}{255, 255, 192}
\definecolor{pastelpink}{RGB}{255, 192, 203}
\definecolor{pastelorange}{RGB}{255, 218, 185}
\definecolor{pastelcyan}{RGB}{173, 216, 230}
\definecolor{pastellavender}{RGB}{230, 230, 250}

\newcommand{\rev}[1]{{#1}}

\begin{document}


\title[Conversate: AI-Mediated Interview Practice]{Conversate: Supporting Reflective Learning in Interview Practice Through Interactive Simulation and Dialogic Feedback}

\author{Taufiq Daryanto}
\affiliation{%
  \institution{Computer Science, Virginia Tech}
  \city{Blacksburg}
  \state{Virginia}
  \country{United States}
}
\email{taufiqhd@vt.edu}

\author{Xiaohan Ding}
\affiliation{%
  \institution{Computer Science, Virginia Tech}
  \city{Blacksburg}
  \state{Virginia}
  \country{United States}
}
\email{xiaohan@vt.edu}

\author{Lance T. Wilhelm}
\affiliation{%
  \institution{Computer Science, Virginia Tech}
  \city{Blacksburg}
  \state{Virginia}
  \country{United States}
}
\email{lancewilhelm@vt.edu}

\author{Sophia Stil}
\affiliation{%
  \institution{Computer Science, Virginia Tech}
  \city{Blacksburg}
  \state{Virginia}
  \country{United States}
}
\email{ssophia@vt.edu}

\author{Kirk McInnis Knutsen}
\affiliation{%
  \institution{Computer Science, Virginia Tech}
  \city{Blacksburg}
  \state{Virginia}
  \country{United States}
}
\email{kirkk@vt.edu}

\author{Eugenia H. Rho}
\affiliation{%
  \institution{Computer Science, Virginia Tech}
  \city{Blacksburg}
  \state{Virginia}
  \country{United States}
}
\email{eugenia@vt.edu}

\renewcommand{\shortauthors}{}

\begin{abstract}
  Job interviews play a critical role in shaping one's career, yet practicing interview skills can be challenging, especially without access to human coaches or peers for feedback. Recent advancements in large language models (LLMs) present an opportunity to enhance the interview practice experience. Yet, little research has explored the effectiveness and user perceptions of such systems or the benefits and challenges of using LLMs for interview practice. Furthermore, while prior work and recent commercial tools have demonstrated the potential of AI to assist with interview practice, they often deliver one-way feedback, where users only receive information about their performance. By contrast, \textit{dialogic feedback}, a concept developed in learning sciences, is a two-way interaction feedback process that allows users to further engage with and learn from the provided feedback through interactive dialogue. This paper introduces \textsc{Conversate}, a web-based application that supports reflective learning in job interview practice by leveraging large language models (LLMs) for interactive interview simulations and dialogic feedback. To start the interview session, the user provides the title of a job position (e.g., entry-level software engineer) in the system. Then, our system will initialize the LLM agent to start the interview simulation by asking the user an opening interview question and following up with questions carefully adapted to subsequent user responses. After the interview session, our back-end LLM framework will then analyze the user's responses and highlight areas for improvement. Users can then annotate the transcript by selecting specific sections and writing self-reflections. Finally, the user can interact with the system for dialogic feedback, conversing with the LLM agent to learn from and iteratively refine their answers based on the agent's guidance. To evaluate \textsc{Conversate}, we conducted a user study with 19 participants to understand their perceptions of using LLM-supported interview simulation and dialogic feedback. Our findings show that participants valued the adaptive follow-up questions from LLMs, as they enhanced the realism of interview simulations and encouraged deeper thinking. Participants also appreciated the AI-assisted annotation, as it reduced their cognitive burden and mitigated excessive self-criticism in their own evaluation of their interview performance. Moreover, participants found the LLM-supported dialogic feedback to be beneficial, as it promoted personalized and continuous learning, reduced feelings of judgment, and allowed them to express disagreement.

\end{abstract}



\begin{teaserfigure}
   \includegraphics[width=\textwidth]{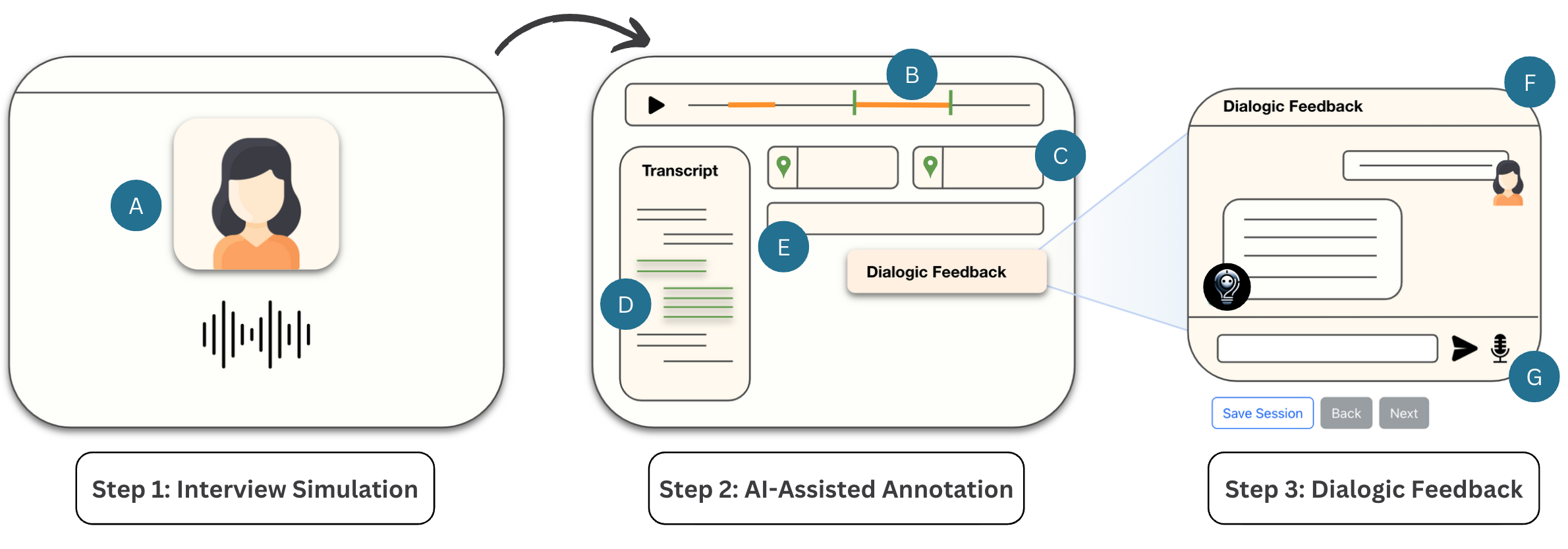}  
    \caption{The overview of \textsc{Conversate} supporting reflective learning in interview practice. \textbf{Step 1: Interview Simulation}. The AI agent conducts an interactive interview simulation, asking questions and dynamically generating contextually relevant follow-up questions based on the user’s responses using an LLM. \textbf{Step 2: AI-Assisted Annotation}. After the simulation, the system analyzes the user's responses and hints suggested areas for improvement by highlighting the corresponding timestamps on the playback bar. (B). Users can then select specific transcript sections (C, D), either based on their own assessment or the LLM-highlighted areas, and provide self-reflections (E). \textbf{Step 3: Dialogic Feedback}. The user can then engage in dialogic feedback (F) focused on the annotated moments, learning from and iteratively refining their answers (G) based on the feedback from the AI agent.}
    \label{fig:teaser}
\end{teaserfigure}

\received{May 2024}
\received[revised]{August 2024}
\received[accepted]{October 2024}


\maketitle

\section{Introduction}
Job interviews are a critical first step in shaping an individual's career trajectory. Research shows that interview performance is a significant determinant of employment offers and subsequent work success, which plays a crucial role in hiring decisions \cite{podsakoff_EffectsOrganizationalCitizenship_2011, mcdaniel_ValidityEmploymentInterviews_, goldberg_WhoGetsHired_1998}. \rev{Performing well in} job interviews, however, is not without its challenges for many people. Many individuals struggle with low confidence \cite{dimopoulos_ApplicantSelfConfidence_2020}, anxiety \cite{mccarthy_MeasuringJobInterview_2004, feeney2015applicant}, and imposter syndrome \cite{posthuma_EmploymentInterviewValidity_2002} despite having the required qualifications for a job position. Such obstacles often diminish the candidate's ability to convey their abilities for the position, leading to poor interview performances \cite{feiler_BehavioralExpressionJob_2016}. 

Research has shown that mock interviews can help alleviate some of these barriers by providing an opportunity for individuals to practice answering interview questions and receive feedback with the help of a practice interviewer \cite{huss_MockInterviewsWorkplace_2017}. The process typically involves simulating real interview scenarios, where the practice interviewer acts as the employer, asking interview questions and providing critique based on the responses \cite{hansen2009best}. This helps most interviewees to learn and assess their strengths and areas for improvement in a lower-pressure environment, allowing them to improve their performance prior to an actual interview \cite{huss_MockInterviewsWorkplace_2017}. In fact, engaging in mock interviews is generally associated with higher confidence and a greater likelihood of success in real interviews \cite{maurer_RelationshipCoachingPerformance_1998, maurer_IntervieweeCoachingPreparation_2001} across most industry sectors. 

\rev{Yet} despite such benefits, mock interview settings, which typically require another person as a practice interviewer, can have limitations. For instance, not all individuals have access to career counselors or expert interviewers \cite{schaub_RolePersonalityLearning_2005}, especially in underserved communities and rural areas where professional staff may be limited \cite{mckenzie2007attracting}. Furthermore, mock interviews can take up a substantive amount of time \cite{marks2006round}.  Logistical issues\rev{, such as scheduling conflicts,} also restrict access to qualified mock interviewers \cite{MACH}. In addition, the cost of hiring mock interviewers could be very expensive. Platforms like Blind \cite{teamblind} or Interviewing.Io \cite{interviewingio} allow individuals to connect with current employers at top technology companies like Amazon, Google, and Facebook with whom they can engage in mock interview sessions. These sessions, however, can cost anywhere from \$100 per hour to \$1800 per package, which consists of several practice interview sessions over time. A more financially viable option would be to practice with friends or peers. However, research shows that practicing with peers can be associated with discomfort due to personality differences \cite{ caldwell_PERSONALITYCHARACTERISTICSJOB_1998}.

To address some of these challenges, both industry and academia have begun to explore the potential of leveraging artificial intelligence (AI), particularly large language models (LLMs), to facilitate mock interviews \cite{MACH, TARDIS, inoue2020jobErica2, rasipuram2020automatic, NaimAnalysisInterview, Chou2022-mm, thakkar2023automatic, li2023ezinterviewer}. For instance, tech companies have developed interview practice systems, such as Meta.AI Mock Interview \cite{metaai}, FinalRoundAI \cite{finalRoundAI}, and Yoodli \cite{yoodli}, which are supported by LLMs that facilitate interview simulations and provide feedback.

Nevertheless, LLM-mediated interview practice systems offer great potential to overcome some of the traditional barriers in mock interviews with human partners, and are gradually amassing public excitement and anticipation \cite{atlas2023chatgpt}. However, our knowledge of how users interact with or perceive such systems remains limited. Furthermore, a key limitation of commercially available LLM-mediated mock interview systems is that their feedback is unidirectional - meaning the system outputs feedback to the user based on the user's performance without further opportunity for in-depth engagement as part of the user's learning process. For instance, Yoodli.AI \cite{yoodli} gives users a general assessment of the user's performance with suggested improvement strategies based on the user's responses to the questions. Although these insights are helpful in supporting lower levels of reflection in learning \cite{fleck2010reflecting}, researchers argue that feedback should actively involve the user in the learning process rather than just delivering assessments \cite{Boud2015-ix}.


To make learning through feedback more interactive and meaningful to the user, some studies have developed and investigated other feedback approaches, such as the \textit{dialogic feedback} method \cite{steen2017dialogic, nicol2014monologue, sutton2009towards, carless2016feedback, carless2012trust}. Dialogic feedback refers to learning about and from feedback through dialogue by actively involving learners in the sense-making and interpreting of the provided feedback \cite{steen2017dialogic}. Dialogic feedback incorporates two-way interaction in the feedback process \cite{carless2016feedback} where the learner is not a mere recipient of feedback but an active participant who intentionally seeks feedback for learning by engaging in dialogue with the feedback provider \cite{Boud2015-ix}. Studies have shown that this interactive approach allows users to engage more deeply with the provided feedback, leading to more efficient up-skilling \cite{djalilova2019dialogic} and better learning experiences and outcomes \cite{hill2019improving}. 

Some studies in Human-Computer Interaction (HCI) have demonstrated the benefits of using interactive discourse mediated by LLM agents to improve users' learning experiences \cite{chopra2024exploring, ross2023programmer}. Such studies are in line with the idea of dialogic feedback, which highlights the importance of dialogue as a means of encouraging users to actively engage with the feedback process.  Currently, most commercial mock interview systems do not adopt such approaches. While the nature of open-domain conversation capabilities in LLMs theoretically allows users to inquire further about their performance, the provided feedback is often prescriptive \cite{yoodli}, limiting the extent to which users can take a proactive learning approach \cite{Boud2015-ix} and engage in agentic sense-making \cite{steen2019students} from the feedback process.

To address this gap, we introduce \textsc{Conversate}, an LLM-supported web application that facilitates job interview practice with dialogic feedback. \textsc{Conversate} incoporates three main steps: 1) \textbf{Interview Simulation}. An AI agent conducts an interactive interview simulation, asking initial questions (e.g., "Tell me about yourself?") and dynamically generating contextually relevant follow-up questions based on the user's responses using an LLM. 2) \textbf{AI-Assisted Annotation}. After the mock interview session, the system analyzes the user's audio-recorded responses and hints suggested areas for improvement by highlighting the corresponding timestamps on the playback bar. Users can also identify potential improvement areas by selecting timestamps corresponding to responses they felt were weak. As part of their annotations on the selected timestamps, users can write self-assessment notes, or engage with the LLM agent by asking questions or requesting targeted feedback on the selected areas. 3) \textbf{Dialogic Feedback}. Users can then engage in interactive discussions around the feedback provided by the LLM agent based on their annotations. For instance, users can ask the LLM-agent to \rev{provide further feedback, clarify its feedback,} or request examples that demonstrate the feedback by conversing with the LLM agent. After this interactive sense-making discussion, the user can then immediately apply their refined understanding of the feedback by revising their responses, allowing them to not only put their learning into practice right away, but also to iteratively improve their skills through repeated practice. 


To evaluate \textsc{Conversate}, we conducted a user study with 19 participants to learn about their perception and use of our system. Our findings show that one major benefit of practicing with an AI as opposed to a human partner was the affordance to make mistakes in a low-stakes environment for repeated practice. Participants also found the AI-assisted annotation particularly helpful in reducing their cognitive burden by prioritizing potential areas that needed improvement. Moreover, dialogic feedback allows participants to engage in continuous back-and-forth interaction with the AI, allowing them to engage with the feedback actively, seek to understand areas for improvement, and promote continuous learning. Additionally, participants valued the ability to revise their answers while receiving dialogic feedback through the system's revision feature, as it enabled them to apply the feedback immediately. Furthermore, our study also discusses the necessity to balance low-stakes AI practice with high-pressure realism. Moreover, while LLM-based dialogic feedback has its potential, several challenges exist; for example, when users express disagreement during the dialogic feedback, the LLM easily agrees with users' disagreement, which is related to the sycophantic behavior of LLMs \rev{(i.e., LLMs tend to respond to user by providing the answers that align with the user's preferences \cite{perez2022discoveringAnthropic})}.

The key contributions of this work are as follows:
\begin{itemize}
    \item \textsc{Conversate}, an interactive web-based application that leverages LLMs to facilitate reflective learning through interactive interview simulation, AI-assisted annotation, and dialogic feedback for job interview practice.
    \item Insights into the users' perceptions of using LLM-powered conversational AI for interview practice.
    \item Discussions on the benefits of dialogic feedback, comparison of LLM-based dialogic feedback with dialogic feedback from humans, challenges of using LLMs for dialogic feedback, and implications of dialogic feedback.
\end{itemize}

\section{Background}
\subsection{Technology-Assisted Interview Practice}
Practicing for job interviews is important for job seekers to improve their performance and reduce anxiety during actual interviews \cite{williams2008effects}. One practical approach to enhancing interview readiness is interview simulations \cite{reddan2008benefits}. Early virtual agents for interview simulation, such as MACH \cite{MACH} and TARDIS \cite{TARDIS}, focused on asking common questions and recognizing user responses through nonverbal cues. However, they lacked the ability to actively listen and respond appropriately to the conversation context, which is essential for effective interview practice \cite{rogers1984active, weger2010active}. To enhance interactivity, recent studies have enabled agents to ask follow-up questions by leveraging machine learning models. These models classify responses and generate relevant questions based on the user's preceding response \cite{inoue2020jobErica2, rao2021FollowUpQuestionGenerator}, consider dialog history, resumes, and job descriptions \cite{li2023ezinterviewer}, and external knowledge from the web \cite{DuerQues}. Notably, studies demonstrate that practicing with such virtual agents can be as effective as practicing with human interviewers \cite{damian2015games, lucas2014s, lucas2017reporting}, suggesting their potential as valuable interview preparation tools.

Researchers have also studied various mechanisms for providing feedback following practice sessions. For example, Chou et al. \cite{Chou2022-mm} evaluated the interviewees' verbal responses from the context of soft skills, while Naim et al. \cite{NaimAnalysisInterview} evaluated their responses on fluency and structure. Advancements in multimodal deep learning have also allowed researchers to develop systems that can evaluate users' non-verbal communication skills during interview sessions \cite{thakkar2023automatic}. 
Similarly, recent advancements in LLMs and conversational AI like ChatGPT \cite{openai2023gpt4} enable more human-like dialogue capabilities, which presents an opportunity to enhance the interview practice experience by facilitating natural conversations akin to actual interviews \cite{chatGPTforEd}. Commercial tools like Meta.AI Mock Interview \cite{metaai}, FinalRoundAI \cite{finalRoundAI}, and Yoodli \cite{yoodli} have leveraged LLMs to simulate interactive mock interviews. While promising, how users perceive such LLM-powered AI systems for interactive interview practice is unclear. Building upon prior studies and recent technological developments in mock interview systems, we aim to examine how users perceive and interact with  LLM-supported systems for mock interview practice and feedback.

\subsection{Facilitating Reflective Learning Through Technological Design} \label{related_work:reflective_learning}
The concept of reflective learning refers to critically examining self-experiences to gain a deeper understanding and learn from them \cite{boud2013reflection}. By doing so, reflective learning helps users to identify areas for improvement and develop their skills. In the context of interview practice, reflection is important for individuals to think about their experience and how they respond to interview questions \cite{lackner2017helping}. As such, studies have demonstrated how incorporating reflective learning in mock interviews aids not only people's learning experience and process but also their interview skills and performance \cite{lackner2017helping}. 

Prior HCI researchers have incorporated various technological designs in training systems to support reflective learning \cite{chen2023facilitating, arakawa2020inward, bridgeman2024using, zhou2021virtual}.  Some of these features include facilitating role-play \cite{chen2023facilitating},  reviewing recordings \cite{arakawa2020inward, zhou2021virtual}, annotation \cite{chen2023facilitating, arakawa2020inward}, and feedback \cite{chen2023facilitating}. For example, Pin-MI \cite{chen2023facilitating} facilitates reflective learning in mental health counseling through role-playing with real-time annotation and peer feedback. Additionally, Zhou et al. \cite{zhou2021virtual} used virtual reality (VR) to enable individuals to review their public speaking recordings, facilitating reflective learning.

Furthermore, designing effective reflective learning systems requires three key components proposed by Slovak et al. \cite{slovak2017reflective}: \textit{explicit}, \textit{social}, and \textit{personal}. The explicit component refers to structuring experiences for learners to simulate their practice, such as through role-playing scenarios or simulations. The social component recognizes the value of a supportive learning environment, for instance, through supportive mentors who can provide feedback. The personal component relies on learners' motivation to actively engage with and learn from their experiences \cite{slovak2017reflective}.

When simulating experiences in reflective learning, Slovak et al. \cite{slovak2017reflective} emphasize that the experience should be realistic enough  \cite{schon1987educating} while ensuring that learners are not overwhelmed by emotions. Hence, to support the \textit{explicit} component of reflective learning, \textsc{Conversate} aims to make interview simulations more realistic by generating contextually relevant follow-up questions. For the \textit{personal} component, users can annotate and self-reflect on their transcripts.
Furthermore, to incorporate the \textit{social} component, \textsc{Conversate} engages users in a dialogic feedback process with the LLM agent.

\subsubsection{Annotation to Support Reflective Learning}
Annotation has been widely used to facilitate reflective learning \cite{chen2023facilitating, arakawa2020inward, mirriahi2018effects, colasante2011using} as it helps people to review and analyze their own recordings \cite{rich2009video}. In the context of interview practice, annotation can be beneficial as it helps individuals recognize important moments and identify their strengths and weaknesses \cite{chen2023facilitating}. However,  annotations during reflection can be overwhelming, especially when the practice session is lengthy \cite{brooks2007surveyExtensive}.
To address this challenge, several studies have explored leveraging machine learning to identify significant moments \cite{arakawa2020inward, arakawa2022human}. For instance, INWARD \cite{arakawa2020inward} uses machine learning to detect important visual cues during video-recorded executive coaching sessions, assisting users in annotating relevant segments for reflective learning. Inspired by these studies, \textsc{Conversate} incorporates an AI-assisted annotation feature. It uses AI to highlight user responses that require improvement, facilitating the annotation process during reflective learning.

\subsubsection{Feedback to Support Reflective Learning}
Feedback is essential in supporting reflective learning in interview practice, as it provides external evaluation and identifies areas for improvement \cite{williams2008effects, anisa2020oral}. 
Several studies in reflective learning have explored the use of feedback to enhance the learning experience, such as incorporating peer feedback \cite{chen2023facilitating, xie2008effect} or AI-generated feedback \cite{huang2023speechmirror}. While peer feedback can be effective by offering new perspectives and facilitating discussions \cite{chen2023facilitating}, it involves the participation of other individuals, which may not be accessible or convenient for individual interview practice \cite{MACH}. Additionally, prior work that provides AI feedback for reflective learning \cite{huang2023speechmirror} typically offers one-way feedback, limiting the ability for users to engage in dialogue and seek clarification or additional guidance. To address these limitations, \textsc{Conversate} aims to implement dialogic feedback to support reflective learning in interview practice.

\subsection{Dialogic Feedback} \label{related_work:dialogic_feedback}
Feedback is important to enhance interview performance as it identifies areas of improvement \cite{Hooda2022-sw}. However, feedback involves more than simply providing information to learners about their practice \cite{Boud2015-ix}. Effective feedback necessitates engaging users in a way that results in noticeable changes \cite{Boud2015-ix}. Prior work by Boud et al. \cite{Boud2015-ix} emphasizes that for feedback to be effective, the feedback loop must be closed, meaning that feedback is not only provided but also received, understood, and acted upon by the learners. Such studies emphasize that meaningful engagement with feedback for learners is a \textit{process} rather than an outcome, where learners are not merely recipients of information but active agents who actively seek information for learning. One way to achieve this is through a two-way interaction between the learner and the feedback provider, where the learner plays a more active role in seeking and using the feedback \cite{Boud2015-ix}. This two-way interaction in the feedback process can be termed "Dialogic Feedback."

Several studies have focused on conceptualizing and exploring dialogic feedback \cite{steen2017dialogic, nicol2014monologue, sutton2009towards, carless2016feedback, carless2012trust}. Based on Steen et al. \cite{steen2017dialogic}, dialogic feedback refers to learning about and from feedback through dialogue, involving learners in actively interpreting and making meaning from the feedback. It is a process where feedback occurs through a conversational exchange, enabling learners to seek clarification, ask questions, and engage with the feedback to construct a deeper understanding \cite{steen2017dialogic}.

To conceptualize dialogic feedback, Steen et al. \cite{steen2017dialogic} suggested four main dimensions that are important to support learners: (1) emotional and relational support, where dialogic feedback is viewed as a relational and emotional process, thus facilitating a supportive learning environment using empathetic language is important; (2) maintenance of the dialogue, ensuring the continuation of the feedback dialogue to enrich the interactive exchange; (3) opportunities for users to express themselves, allowing learners to articulate their thoughts and reflect on their understandings within the dialogue; and (4) contribution to individual growth, creating opportunities where learners receive feedback as they express themselves, leading to cognitive development.

Similar to the concept of dialogic feedback, some related works in HCI align with the principles of interactive dialogue and feedback exchange aimed at enhancing user's learning experiences. For instance, a study by Chopra et al. \cite{chopra2024exploring} developed an LLM-based conversational AI to assist in software debugging and demonstrated that multi-turn conversational interaction provides users with better explanations than one-way interaction. Another study by Ross et al. \cite{ross2023programmer} showed that users valued an LLM-based conversational programming assistant as it helped them ask follow-up questions and better understand code. \rev{While these prior works have built conversational AI to facilitate learning, they are not grounded in the literature on dialogic feedback theories \cite{steen2017dialogic, nicol2014monologue, sutton2009towards, carless2016feedback, carless2012trust}, meaning that they may overlook key elements that make dialogic feedback more effective for learners (e.g., the four elements that we mentioned above based on Steen et al. \cite{steen2017dialogic}). Hence, most of these studies do not empirically assess the benefits and challenges of incorporating LLM-based dialogic feedback in learning systems. By contrast, the novelty of our work lies in integrating dialogic feedback in an LLM-based learning system by incorporating key components, grounding it firmly in the literature, and empirically examining both the benefits and challenges of leveraging LLMs to facilitate dialogic feedback in a learning context.}

Inspired by prior studies demonstrating the benefits of AI-powered conversations in learning contexts, \textsc{Conversate} incorporates a dialogic feedback approach to facilitate an interactive dialogue between the AI agent and the user. By doing so, we aim to improve the user's meaning-making process by discoursively engaging more in-depth with the provided feedback. Our design implementation is also grounded in the four main dimensions of dialogic feedback identified by Steen et al. \cite{steen2017dialogic}. With this design approach, our work examines how users interact with and perceive LLM-facilitated dialogic feedback in our mock interview system.

\section{Conversate Description}
\rev{There are three main components of \textsc{Conversate}: 1) Interview Simulation, 2) AI-Assisted Annotation, and 3) Dialogic Feedback. These three components align with the key components in the reflective learning system proposed by Slovak et al. \cite{slovak2017reflective}: explicit, personal, and social (\S \ref{related_work:reflective_learning}). \textsc{Conversate} supports the explicit component of reflective learning through the interactive Interview Simulation. For the personal component, users can annotate and self-reflect on their transcripts using the AI-assisted annotation feature. Furthermore, to incorporate the social component, Conversate engages users in a Dialogic Feedback process with the LLM agent. The design decision for each component was informed by existing literature, as explained in the related work section and described further in each subsection. }

\subsection{User Flow Overview}
When using \textsc{Conversate}, the user flow begins with the user entering the job title they want to practice an interview for. They then engage in an \textbf{Interview Simulation} session with an AI agent. The AI agent asks pre-defined interview questions one by one. After each question, the user records their verbal response, and the AI agent dynamically generates and asks a contextually relevant follow-up question based on the user's response. This back-and-forth interaction continues until the user has answered all the questions.

Upon completing the interview simulation session, the user proceeds to the \textbf{AI-Assisted Annotation} step. In this step, the system analyzes the user's recorded responses and highlights areas for improvement. The user can then view the transcript, select specific sections based on the highlighted areas or their own preferences, and provide self-reflections on their performance.

Next, the user engages in \textbf{Dialogic Feedback} focused on the annotated moments. The user interacts with the dialogic feedback to help them learn from the feedback and iteratively refine their answers. Once this process is finished, the user can save this session and get feedback on other moments as desired.

\subsection{Interview Simulation}
The first step provides an interactive simulation for users to practice answering interview questions with the help of an AI agent. The AI agent verbally asks interview questions, and the user records their answers. Leveraging GPT-3.5-Turbo from OpenAI, combined with their text-to-speech and speech-to-text models, the AI agent can dynamically generate and ask contextually relevant follow-up questions based on the user's responses. This aims to enhance the interactivity and realism of the simulated interview experience.

\begin{figure*}[t]  
    \centering
    \includegraphics[width=0.75\textwidth]{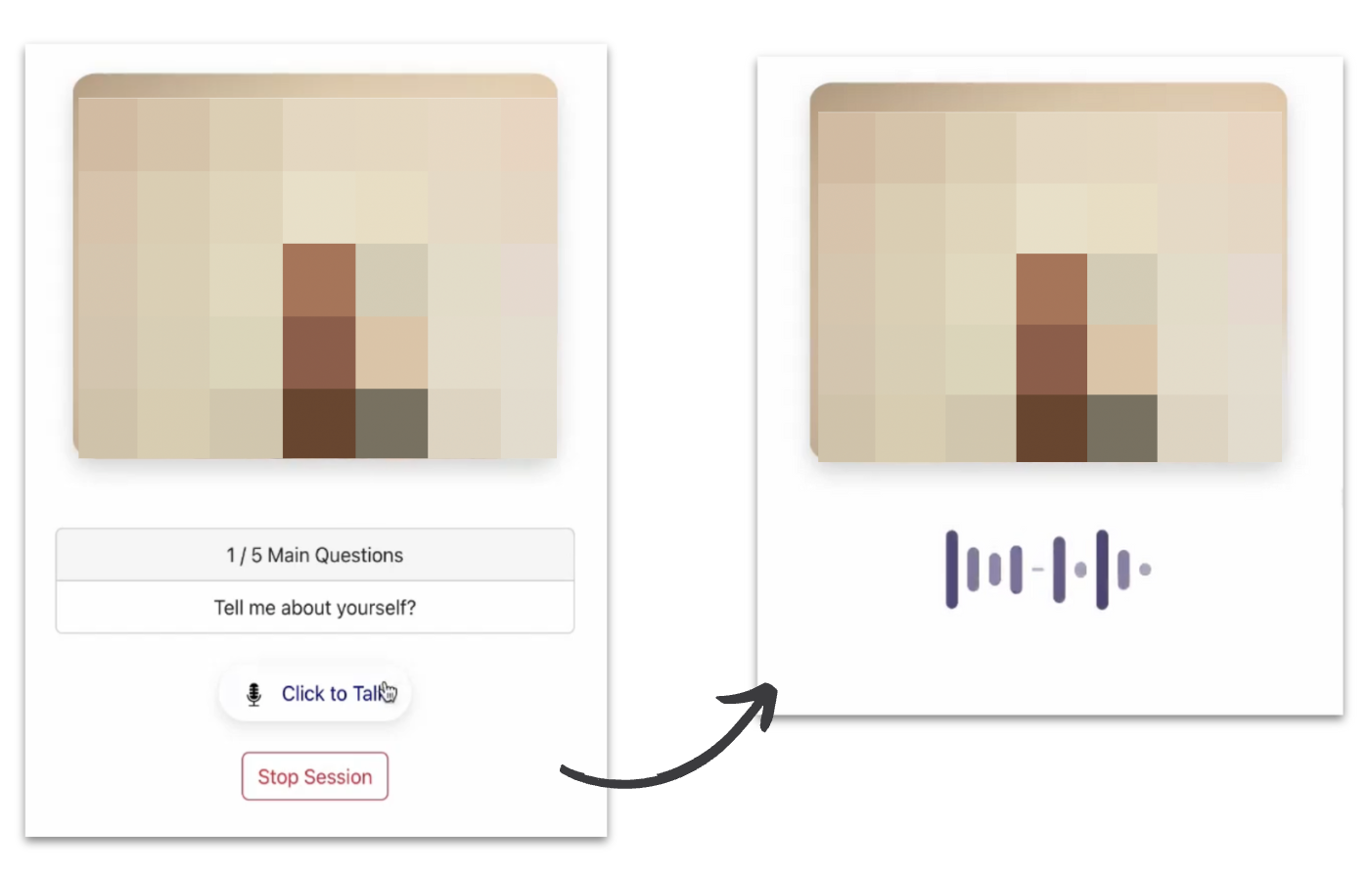}  
    \caption{\textbf{Step 1: Interview Simulation}. \rev{The AI agent conducts an interactive interview simulation, asking initial questions (e.g., "Tell me about yourself?") and dynamically generating contextually relevant follow-up questions based on the user’s responses using an LLM. \textit{Note: The pixelated area shows the participants who appeared on camera. It has been pixelated to comply with anonymization rules.}}}
    \label{fig:interview_simulation}
\end{figure*}

For the interview simulation, we curated a primary set of common interview questions drawing from prior research \cite{akhtar2012expect} and consultations with two career counselors and a career center director at an R1 university. In the simulation, we prompted GPT-3.5-Turbo to act as the interviewer agent, asking questions one by one to the user. In response to each question, the user records their response. Subsequently, GPT-3.5-Turbo generates a contextually relevant follow-up question that is adapted based on the user's previous responses and conversation history. This approach aims to make the interview simulation more realistic and aligned with real-world interview settings, as follow-up questions from the interviewer in actual interviews are commonly based on interviewee responses  \cite{levashina2014structured}. During the user study, we restricted the number of follow-up questions for each main question to one in order to maintain a manageable duration for the simulated interview session. 


\subsection{AI-Assisted Annotation}
\begin{figure*}[t]  
    \centering
    \includegraphics[width=\textwidth]{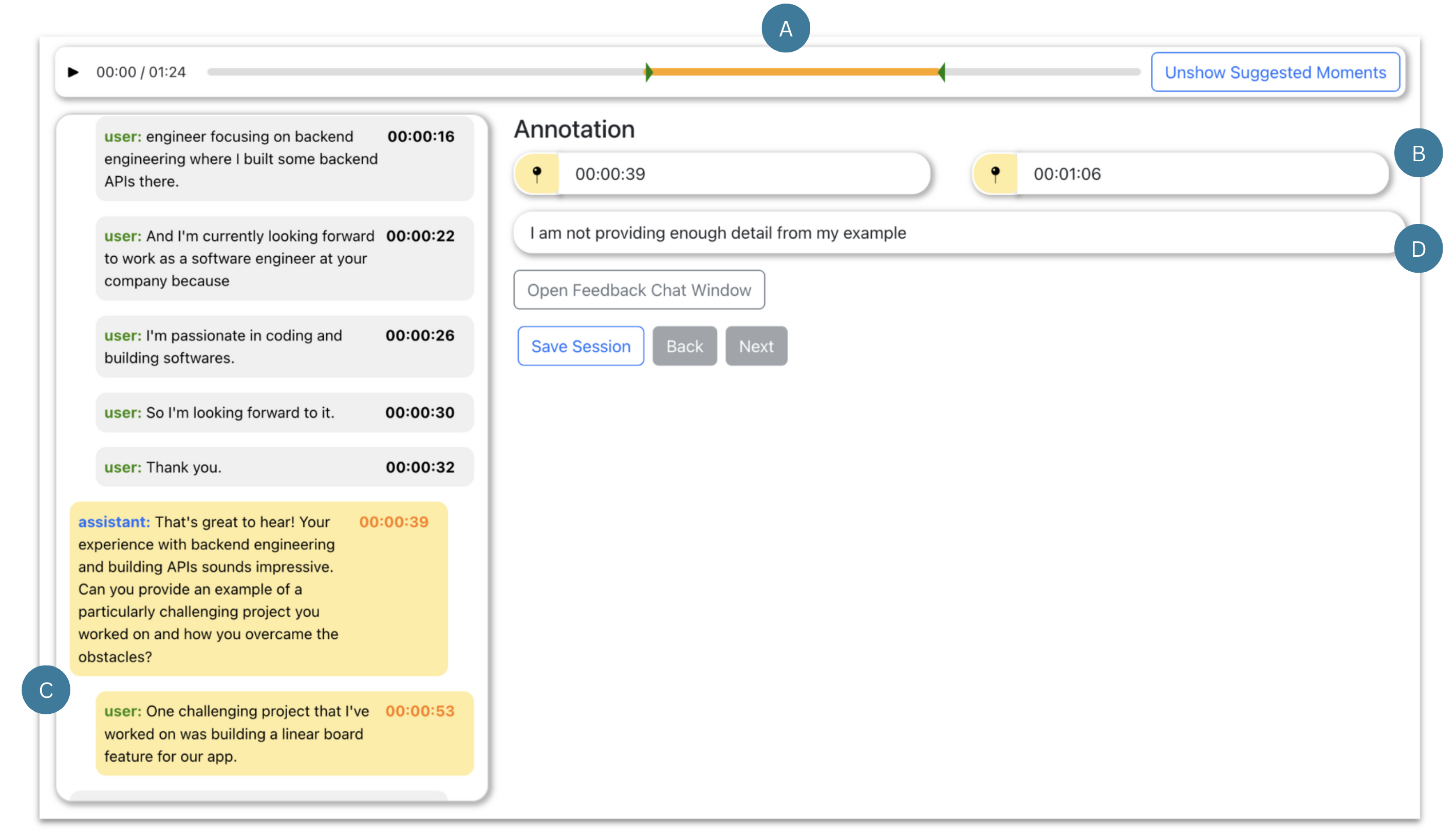}  
    \caption{\textbf{Step 2: AI-Assisted Annotation}. After the simulation, the system analyzes the user’s responses and highlights areas for improvement (A). Users can then select specific transcript sections (B, C), either based on their own preferences or the highlighted areas, and provide self-reflections (D).}
    \label{fig:ai_assisted_annotation}
\end{figure*}

\textbf{Transcript-Navigation \& Playback Bar}: After the interview session, our system will show a timestamped transcription of the interview as shown in the left-hand panel of  (Fig \ref{fig:ai_assisted_annotation}). Additionally, we implement a transcript-based navigation feature to support user interaction, allowing users to navigate through the audio recording by clicking on sentences in the generated transcript. When a user clicks on a sentence, the audio playback bar on the top will move its cursor to the corresponding timestamp. This enables users to access specific timestamps by reading the transcript and clicking on sentences to hear the corresponding audio recordings. This navigation approach is similar to a feature found on platforms like edX \cite{edX} and TED talk platforms \cite{TED}, which allow users to click on words in the transcript to jump to the specific timestamp \cite{pavel2014video}. 

\subsubsection{AI-Highlighted Hints}
After the interview simulation, the system analyzes the user's recorded responses using GPT-3.5-Turbo and highlights areas as hints to improvement areas in the user's mock interview performance. Specifically, \rev{we created a prompt for the LLM} to evaluate each of the user's answers based on these criteria: answer detail, structure, and adherence to good communication principles, including effectiveness, appropriateness, efficiency, clarity, and competence in answering the interview question correctly \cite{spitzberg2000good}. Answers that lack sufficient detail, have a weak structure, or fail to follow these communication principles are classified by the LLM as needing improvement. These evaluation criteria were developed based on prior research \cite{Chou2022-mm, NaimAnalysisInterview} and in consultation with the director of the career center and two career counselors at the university. We did not do any additional fine-tuning for our model; instead, we used a prompt engineering approach since the focus of this paper is more on the early-stage exploration of user perceptions towards LLM for interview practice (The detailed prompt is written in Appendix \ref{Appendix:Prompt_AI_highlighting}). 

After the LLM evaluates the user's answer, the system then highlights the corresponding sections in the transcript, as well as the relevant time ranges on the audio recording's playback bar in orange, as shown in (Fig. 3A). This highlighting helps users during the annotation process by visually indicating portions of the transcript and audio recording where the AI suggests the user's answers could be refined or expanded upon. By doing so, this feature aims to help users identify and revisit important moments quickly, reducing the cognitive overload of reviewing the entire transcript, which can be time-consuming and tedious \cite{brooks2007surveyExtensive, arakawa2020inward}.

\subsubsection{Annotation}
During this annotation process (Fig \ref{fig:ai_assisted_annotation}), users can select specific time ranges from the recording that they want to reflect on, either based on the highlighted moments or their own preferences. After selecting the relevant time ranges, users can provide comments as part of a self-assessment, reflecting on their performance during those specified moments. After that, they will engage in dialogic feedback focused on the specified moment for further reflection.

This time-range selection also aims to make the following feedback more focused on the specific moments that users wish to analyze. Additionally, the time-range selection also aims to shorten the input context for the LLM, as LLMs often struggle to comprehend and analyze lengthy input contexts, especially when relevant information is in the middle \cite{liu2023lost}.
Directly inputting the entire practice interview transcript and asking the LLM to provide feedback can be problematic due to the potential length of the sessions. By enabling users to select specific time ranges, we aim to improve the quality, relevance of the feedback, and overall user experience by allowing them to focus on specific moments in the interview session for targeted feedback. 

\subsection{Dialogic Feedback} \label{method:dialogic_feedback}

\begin{figure*}[b]  
    \centering
    \includegraphics[width=\textwidth]{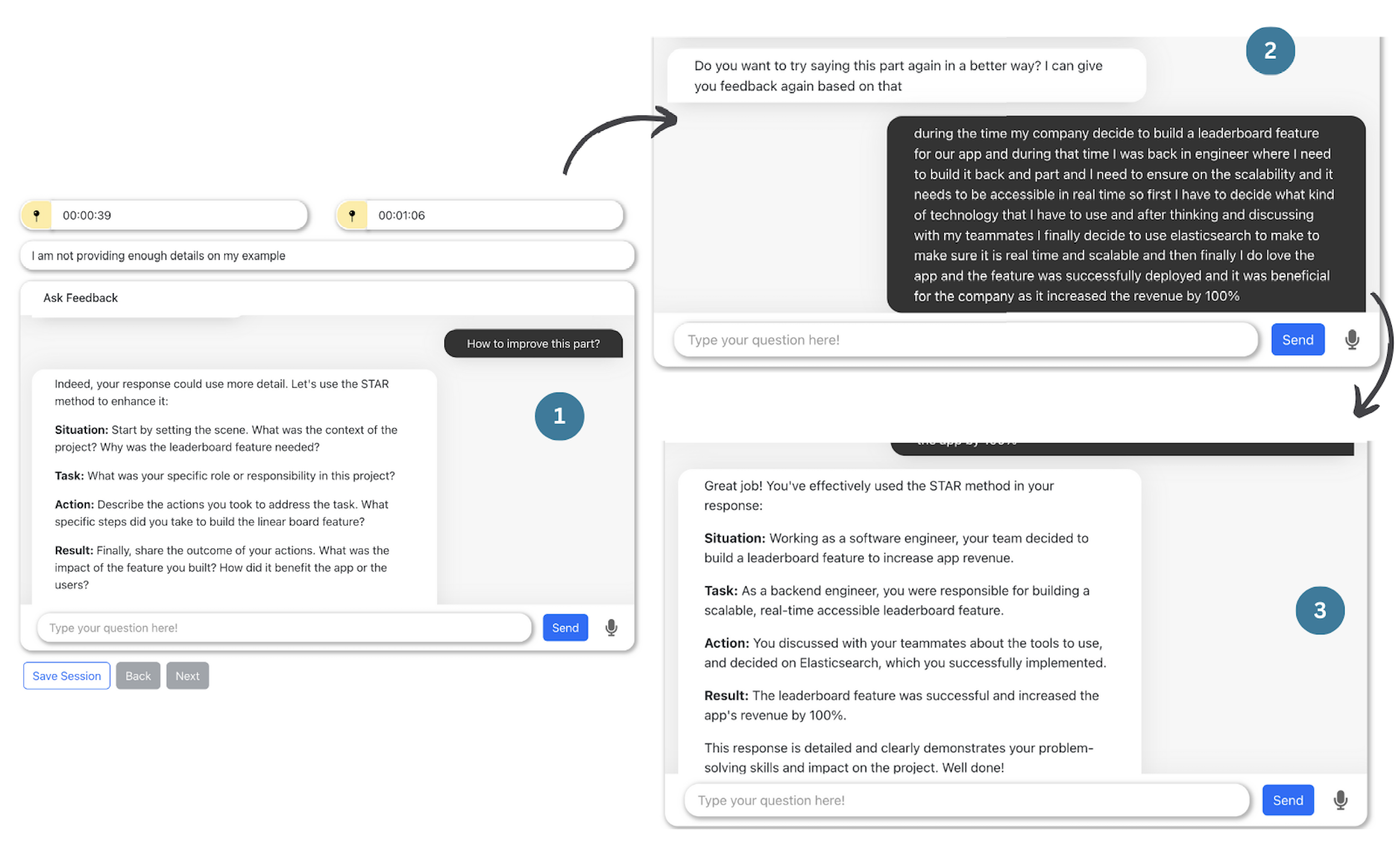}  
    \caption{\textbf{Step 3: Dialogic Feedback}. The user engages in dialogic feedback focused on annotated moments, learning from and iteratively refining their understanding based on the AI agent's guidance. \textbf{(1)} The user initiates the interaction by asking a question (e.g., "How can I improve this part?"), followed by AI feedback. \textbf{(2)} The user refines their answer based on the feedback by clicking the microphone button to record their revised answer verbally. \textbf{(3)} The user receives further feedback or affirmation from the AI. This iterative process of refining and receiving feedback can continue if the user's answer is not satisfactory yet or if the user wants to improve their answer further.}
    \label{fig:scheme_dialogic_feedback}
\end{figure*}

After selecting specific time ranges, users engage in dialogic feedback (Fig \ref{fig:scheme_dialogic_feedback}) with the AI agent powered by GPT-4. This interaction begins with users asking questions (e.g., 'How can I improve my answer?'). The AI agent provides feedback by considering the user's transcript from the specified time ranges selected by the user or the areas highlighted by the LLM agent. This dialogic feedback takes the form of a chat-based interaction, allowing users to ask questions and engage in discussion. \rev{A complete example of a dialogic feedback interaction, based on a sample annotated transcript, can be found in the Appendix \ref{Appendix:Dialogic_feedback_interaction}.}

This dialogic feedback feature is designed to provide a more interactive way for users to participate in their feedback-learning process, thereby improving their learning experience \cite{Boud2015-ix, carless2016feedback}. This approach positions the user as an active learner who agentically seeks and learns from feedback through question-answering \cite{Boud2015-ix}. When designing this feature, we consider four main dimensions of dialogic feedback (\S \ref{related_work:dialogic_feedback}) to enhance learning \cite{steen2017dialogic}:

\begin{enumerate}
    \item Emotional and relational support: The LLM is prompted to acknowledge the user's answer in a supportive manner by offering constructive feedback to enhance their responses. This approach aims to boost learners' self-esteem \cite{nicol2006SevenFeedbackPrinciple} and to avoid negative self-perceptions \cite{bentvelzen2022revisiting}. 
    \item Maintenance of the dialogue: Upon receiving feedback, users can ask additional questions to deepen their understanding of the LLM agent's feedback. The LLM agent will then further engage by responding to the user, thereby maintaining a continuous dialogue with the user.
    \item Opportunities for users to express themselves: During the dialogic feedback process, we added a \textit{\textbf{revision feature }}for users to revise their answers by clicking the record button and revising their responses post-feedback. This is to provide users a chance to immediately apply their understanding of the feedback into action \cite{Boud2015-ix}. 
    \item Contribution to individual growth: After users revise their answers, the LLM will provide feedback again, and users can continue revising repeatedly. This iterative process offers users multiple opportunities to improve their answers as repeated practices have been significantly associated with improved learning up-take and outcomes \cite{bruner2001repetition, jung2013effectTaskRepetion}.
\end{enumerate}


Additionally, we designed our back-end prompting framework such that the LLM-agent provides feedback grounded in communication theories \cite{spitzberg2000good} and the STAR (Situation, Task, Action, Result) method \cite{Boyer}. The STAR method is one of the most widely used frameworks for answering job interview questions \cite{akhtar2012expect}, where candidates break down their responses into distinct components: the situation they faced, the task at hand, the specific actions they took, and the resulting outcomes. By offering a step-by-step breakdown with examples, we align with the effective feedback principle of clarifying what constitutes good performance \cite{nicol2006SevenFeedbackPrinciple}. This approach helps learners understand the criteria for a well-structured, comprehensive answer.

\subsection{Implementation Notes}
The \textsc{Conversate} web application is implemented using HTML, CSS, Bootstrap, and TypeScript, with the VueJS framework \cite{vuejs} used for building UI components. The backend uses Python with Flask framework \cite{flask}. It incorporates OpenAI's text-to-speech model "tts-1" \cite{openaitts} for converting interview questions into voice and uses the "whisper-1" speech-to-text model \cite{openaitts} for transcribing user voice recordings into text. For simulating interviews, OpenAI's GPT-3.5-Turbo \cite{gptmodel} is used to generate follow-up questions based on user responses. Additionally, ``GPT-4-0613`` \cite{gptmodel} is used to provide dialogic feedback. User data is stored in MongoDB \cite{mongodbatlasdatabase} for textual data and Firebase cloud storage \cite{firebasestorage} for voice recordings. The production version of the web application is hosted on Vercel \cite{vercel}.

\section{User Study}
\subsection{Participants}
Nineteen participants were recruited through various channels, including university announcements, online advertisements (Craigslist), social media platforms (Reddit and Facebook groups), and word of mouth. During the recruitment process, we initially distributed a screening questionnaire containing inquiries about participants' prior experience with interview practice and demographic information. Subsequently, we selected individuals with prior experience with interview practice while considering demographic diversity. At the end of the user study, participants were compensated with a \$20 Amazon e-gift card. 

Our selected participants were diverse in terms of demographic factors such as age, gender, and occupation. We recruited 19 participants, including 9 university students and 10 non-students (see Table 1). 
All participants were over 18 years old and had experience preparing for job interviews. We also included participants who had not attended any job interviews yet as long as they had experience preparing for one.

\begin{table}[b]
\centering
\caption{Participant Demographics}
\begin{tabular}{|c|c|c|c|c|c|}
\hline
ID & Gender & Age & Occupation & Interview Count & Practice Frequency\\
\hline
P1 & Male & 26 & PhD student & 1-5 & Occasionally \\
P2 & Other & 25 & Receptionist & 1-5 & Occasionally \\
P3 & Male & 30 & Social worker & 6-10 & At least once a week \\
P4 & Male & 25 & Clerk at a store & 1-5 & Everyday \\
P5 & Male & 22 & Undergraduate student & 1-5 & At least once a week \\
P6 & Female & 28 & Undergraduate student & 0 & At least once a week \\
P7 & Male & 25 & Undergraduate student & 1-5 & Everyday \\
P8 & Male & 21 & Undergraduate student & 1-5 & At least once a week \\
P9 & Male & 30 & Master student & >10 & Occasionally \\
P10 & Female & 19 & Undergraduate student & 1-5 & At least once a week \\
P11 & Female & 25 & Unemployed & >10 & Everyday \\
P12 & Female & 23 & Paralegal at law company & 6-10 & Everyday \\
P13 & Male & 20 & Undergraduate student & 1-5 & Everyday \\
P14 & Female & 34 & Unemployed & >10 & Occasionally \\
P15 & Female & 31 & Unemployed & 6-10 & Occasionally \\
P16 & Female & 21 & Undergraduate student & 1-5 & Occasionally \\
P17 & Female & 33 & Operations specialist & >10 & Everyday \\
P18 & Male & 40 & Government consultant & >10 & Everyday \\
P19 & Male & 23 & Unemployed & 1-5& Occasionally \\
\hline
\end{tabular}
\label{tab:participants}
\end{table}

\subsection{Procedure}
We conducted a semi-structured interview with the participants. First, we explained the \textsc{Conversate} tool and showed the participants a demo of the tool. We then asked the participants to engage with the tool by using the interview simulation, AI-assisted annotation, and dialogic feedback. Participants interacted with the AI agent by answering four main interview questions, each followed by a follow-up question, for approximately 10 minutes \cite{chen2023facilitating}. Subsequently, they used the AI-assisted annotation and dialogic feedback features within a 15-minute timeframe while using our system. We then followed up with questions to gather insights into their perception and experience with our system. During this session, we also asked questions about their prior experiences in preparing for and practicing job interviews and using AI assistance for interview practice.

\subsection{Data Collection}
\textit{Audio-Video Recordings \& Researcher Notes}: Participants were recorded when using the tool. The researchers additionally made observational notes during the sessions.

\textit{User Input}: While participants were using the tool, all their user inputs, such as the interaction between participants and the AI during the practice, the moments they selected during the annotation, and their interactions with the AI in dialogic feedback, were collected and stored in the database. 

\textit{Interview}: After participants completed using the tool, the researchers conducted an audio-recorded semi-structured interview with them. The interview included semi-structured questions and additional questions pertaining to the observations made by the researchers during the participants' use of the tool. We also asked participants to recall notable moments when interacting with our system. Through this approach, we were able to identify specific interactions that were mentioned during the interview.

\subsection{Data Analysis}
The interview data was transcribed using Zoom, and then a researcher manually verified the accuracy of the transcript against the original recording. We used a thematic analysis \cite{braun2012thematic} approach to analyze the interviews. The first author conducted an initial round of open coding \cite{charmaz2006constructing} on the interview transcripts. Subsequently, this researcher presented the open codes to the three other researchers for discussion and finalization. Afterward, the researcher used the finalized list of codes to code the interviews.

We also complemented our qualitative analysis by observing user interactions with the AI based on the user interaction data logged on our server. Specifically, we noted how participants used the AI feedback, including the questions they asked the AI, the AI's responses, and the revisions made to the users' answers.

\section{Findings}
\subsection{Perceived Benefits of Interview Practice Simulation With AI}

\subsubsection{Fostering Improvement Through Repeated Practice in a Low-Stakes Environment.} \label{finding:AI_simulation_1}
One major benefit of practicing with an AI as opposed to a human partner was the affordance to make mistakes in a low-stakes environment for repeated practice. For example, according to P17, \textsc{Conversate} allows you to “make your mistakes behind the scenes." Additionally, for many like P6, blundering in front of a computer as opposed to a human counterpart was less nerve-wracking. Similarly, those with high interview anxiety, such as P7, who worries about “losing my cool in front of an actual person,” perceived AI as a “non-judgmental machine.”  


As such, the absence of a critical human observer is associated with reduced fear of judgment among participants (6 | P1-P2, P6-P8, P17). In return, this allows participants to concentrate solely on formulating their responses rather than managing interpersonal reactions or expectations (3 | P2, P6, P7). P6 emphasized that practicing with AI helped her focus more on "how I thought about those words" during interview practice:

\begin{quote}
\textit{"When I'm practicing with people, I'm just thinking about the words I said rather than how I thought about saying those words. [But] this tool helps me think about how I thought about those words in the first place" - P6}
\end{quote}
By not "feeling nervous," P6 reported that she could delve deeper into her thought process to come up with better responses while interacting with the system. Similarly, according to P7,   practicing with something non-judgmental allows one to "prepare for something judgmental" (P7), namely the real job interview itself.

Furthermore, the low-stakes nature of an AI-facilitated practice interview encouraged participants to engage in repeated practice, easing their fears of judgment (3 | P2, P6, P7). 
P10 perceived such opportunity for repeated practice on specific improvement areas as one of the primary advantages of \textsc{Conversate}, in contrast to human-led sessions where time constraints and the one-take nature of the practice sessions make revision interview responses in-situ nearly impossible.


\subsubsection{Perception of AI as a More Professional Interviewer Than Peers}  \label{finding:AI_simulation_2}
Even though participants felt less pressure when practicing with AI, this does not negate their perception that AI can provide a more formal and serious setting for interview practice: 

\begin{quote}
\textit{"Sometimes I feel like with friends, it can be a little bit more casual, and this kind of [AI] simulates a more serious environment. I actually prefer this a little bit more. [...] this experience itself was, I think, a good practice for interviews." - P10}
\end{quote}

This aligns with how P5 feels when practicing with friends, which is he is "very comfortable with [them]," and therefore does not provide a "professional setting."  By contrast, some participants (P5, P10) were impressed by how professional our LLM agent came across as an interview facilitator. P5 attributed \textsc{Conversate}'s professional demeanor as a mock interviewer to \textit{kinds} of questions it asked and \textit{how} the questions were asked, noting how the system generated "very detailed questions" similar to those asked in real-life interviews for the position P5 was currently seeking. 

\subsubsection{Contextually Relevant Follow-up Questions Encourage Deeper Thinking.}  \label{finding:AI_simulation_3}
A key feature of \textsc{Conversate} that enhances the interview practice is the contextually relevant follow-up questions that are generated from the LLM. Overall, participants mentioned that the follow-up questions were helpful (19 | P1-P19) and made the practice more engaging (8 | P1-P2, P4-P7, P14-P16) and felt more like a real interview (13 | P1-P2, P4-P11, P14-P15, P19). This engagement also helped participants think more deeply about their responses (7 | P1, P4, P6, P9, P13, P15, P17). For instance, P1 felt the follow-ups were able to "test different angles" and prompted him to elaborate more:

\textit{\begin{quote}
\textit{"I think the follow-up questions are pretty good because it's keeping the conversation engaging, and I think the system is asking really good [follow-up] questions based on what I have said, testing different angles as well as picking context from my answer” - P1}
\end{quote}}

\begin{figure*}[t]  
    \centering
    \includegraphics[width=\textwidth]{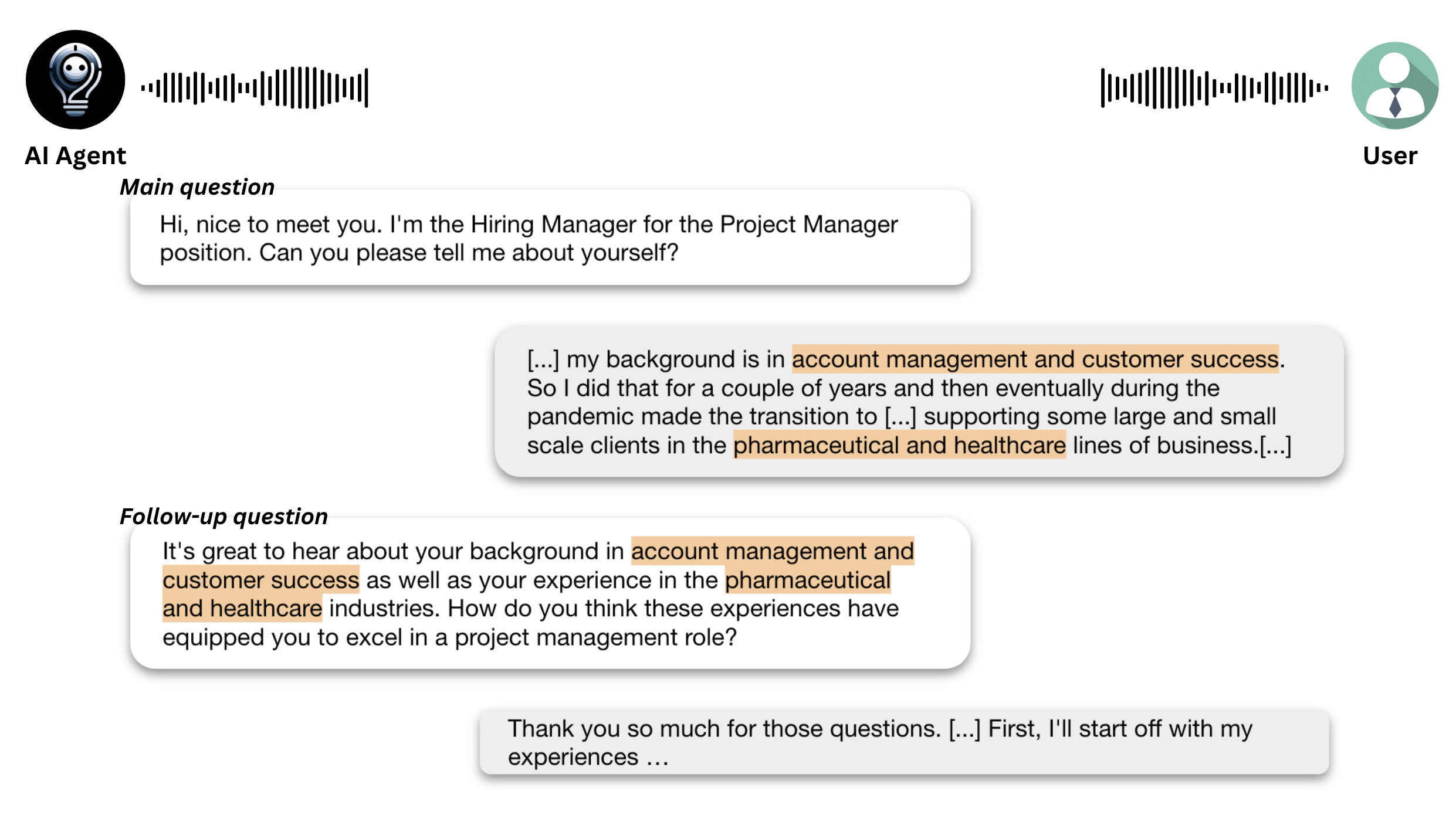}  
    \caption{Sample Conversation During Interview Practice (P9): The text highlighted in orange indicates repeated keywords that made participants feel the follow-up questions were contextually relevant. These follow-up questions encouraged participants to delve deeper into their responses.}
    \label{fig:follow_up}
\end{figure*}

Based on the sample illustration in Fig \ref{fig:follow_up}, we observed that the generated follow-up questions from the LLM contained repeated keywords from the user's response. Those repeated keywords that resonated in the conversation made participants perceive that the follow-up questions were contextually relevant. Additionally, by using follow-up questions, the AI interviewer helped participants like P9 to further elaborate on their initial response. P9 felt the follow-up question "delves further [into my answer] and creates an opportunity for me to explain [more]". 
Additionally, participants found the follow-up questions were valuable, especially when they sometimes struggled to provide good answers, particularly when feeling nervous:
\begin{quote}
\textit{"I know that sometimes, even now, I can get nervous and blank out. So the [follow-up] questions help me think more" - P15}
\end{quote}


For some participants, the follow-up questions posed by our LLM agent nudged them  to "think on the spot": 
\begin{quote}
\textit{"The follow-up questions, in between the main questions, would take part of my answer and then [build on it]. I think that's great practice because the follow-up questions are not usually the kind of things you can find online easily. It felt like it really helped me practice thinking on the spot [about] my real experiences." - P6}
\end{quote}
Some participants expressed that they often felt unprepared when confronted with a question they have not anticipated or practiced in actual interviews (3 | P5, P8, P19).  However, for participants like P6 and P5, the system's follow-up questions encouraged them to think on the spot instead of reciting "responses from your memory" (P5). Furthermore, P13 mentioned that sometimes he does not receive follow-up questions when practicing with human partners.



\subsection{Perceived Benefits of AI-Assisted Annotation}

\subsubsection{Reducing Cognitive Burden by Prioritizing Focus Areas} \label{finding:annotation_1}
Participants found the \textit{AI-Highlighted Hints }feature particularly helpful in reducing their cognitive burden by prioritizing potential areas that needed improvement ( 6 | P1, P6, P8, P11, P12, P16). According to P1, this feature was helpful because reviewing his mock interview session recordings was often overwhelming:

\begin{quote}
\textit{"I think [this feature] is doing a pretty good job because when I clicked on the marked orange bar, it took me to the part of the conversation where I didn't answer very clearly. I think the suggested part was helpful because sometimes it [can be] overwhelming to read [through] a lot of transcript[s]." - P1}
\end{quote}

Similarly, P6 appreciated not having to "hear the entire recording" as she typically did in her own mock interviews with human partners, finding her usual method "more tedious." Overall, participants agreed that the AI-suggested hints allowed them to allocate their learning efforts more efficiently, sparing them from having to review from the very beginning.

Additionally, \textit{AI-Highlighted Hints } helped participants to identify  blindspots , as P17 explained:
\begin{quote}
\textit{"I don't think I would think too heavily about those questions, like with the question 'tell us about yourself.' I gave a very short answer, and I don't think I would have gone back to that specific part unless the orange bar popped up to indicate it." - P17 }
\end{quote}
Previously, P17 did not consider giving a short answer to "Tell me about yourself" as problematic. Hence, the visual cue prompted him to review that part of the conversation. Some participants recalled reviewing and identifying weaknesses in mock interview practice as challenging, especially when practicing alone or with peers. According to P8, people usually only remember "the beginning and the end" of the practice session, leading them to miss other potential areas of improvement.

\subsubsection{Mitigating Excessive Self-Criticism} \label{finding:annotation_2}

Beyond the practical benefits, AI highlighting areas for improvement can also mitigate excessive self-criticism during the reflection. Our study found that some participants (3 | P12, P15, P17) were overly critical of themselves during the reflection and thought their answers were mostly inadequate. P12 mentioned: 
\begin{quote}
\textit{ “I agree with the AI's suggestions about the areas I would need to practice more. [...] Honestly, I probably would have marked more [weak] areas for me. So I think [the AI] is being generous.” - P12  }
\end{quote}
Similarly, P17 felt the "whole bar should be orange,"  indicating she thought her entire interview was poor. These responses suggest the participants were overly harsh on themselves and unable to evaluate their performance objectively. For some, seeing that the AI highlighted only a few areas helped mitigate their tendencies to be overly self-critical (3 | P12, P15, P17)

Interestingly, participants who received numerous highlighted areas did not feel discouraged as long as it helped them improve. As P16 mentioned, despite having many areas marked:
\begin{quote}
\textit{"I don't feel necessarily disappointed because it's just things I need to work on, so I don't feel discouraged at all" - P16}
\end{quote}
The sense of acceptance when receiving suggestions from AI can be partially explained by how participants perceive AI as a non-judgmental agent, providing a safe space to make mistakes in a low-stakes environment (\S \ref{finding:AI_simulation_1}). This also relates to P6's view that AI analyzes answers objectively without any emotional judgment, thus removing the "social fatigue" of receiving feedback from a person:
\begin{quote}
\textit{"When people give feedback, It's often quite weighty, almost emotional, you know? And there's none of that here. They're just analyzing what you said objectively. So I basically don't have the social fatigue of listening to how someone else feels about my answer" - P6}
\end{quote}
Some  hypothesized that receiving extensive highlighted areas could potentially discourage them, as P14 stated:
\begin{quote}
\textit{"If [the AI] says the whole thing was bad, then there's going to be less motivation [for me] to click on every single thing and improve upon it again [...] Even though I think that the whole point is to improve. It is humbling cause we all want to be perfect the first time around [...] I think it needs to have that balance there" - P14}
\end{quote}
According to P14, feeling discouraged may stem from wanting to be "perfect" initially. She perceived that the issue lies not with the AI's suggestions but rather with her internal expectations for flawless performance. Nonetheless, P14 mentioned that there should be a "balance" when the AI provides feedback. If all responses were highlighted as weak areas, some participants might feel discouraged and unmotivated to address everything. Moreover,  this can also negate the practical benefit of this feature to help users focus on specific moments (\S \ref{finding:annotation_1}) if all moments were marked as needing improvement.

\subsection{Perceived Benefits of Dialogic Feedback}
\subsubsection{Dialogic Feedback Promotes Personalized, Continuous Learning}  \label{finding:dialogic_feedback_1}
A major benefit of using dialogic feedback with AI is enabling a personalized learning experience. The ability to "ask your own question" and receive "specific feedback" (P10) made the feedback feel like a "personalized give-and-take dialogue" (P18). According to P6, this dialogue approach allowed her to ask specific questions, which helped overcome "tiny little blockers" hindering her understanding, something generalized feedback alone might not address:

\begin{quote}
\textit{"I think the [dialogue format] is better because if I have any specific or strange questions, the AI can [...] answer that. Usually, I have small little blocks in my brain that  [lead to] weird questions. The AI is good at answering those in dialogue rather than when you get generalized feedback." - P6.}
\end{quote}
This dialogue approach also helps participants keep learning (3 | P13, P17, P18). For instance, P17 mentioned that getting feedback through dialogue allowed her to "continue to ask questions [to] keep advancing your response." Additionally, P18 appreciated this approach as " [it] showed me how to improve, and then gave me a chance to do it, and then it gave me more feedback on that." Participants perceived that continuous back-and-forth interaction allowed them to engage with the feedback actively, seek to understand areas for improvement, and promote continuous learning.

\subsubsection{Dialogic Feedback Helps Users Learn Without Feeling Judged.} \label{finding:dialogic_feedback_2}
The participants' willingness to learn continuously was also attributed to their perception that dialogic feedback allowed them to learn without feeling judged. As P8 stated, engaging in dialogue with the AI "felt less like being evaluated and more like helping you grow." Participants preferred this dialogue approach over being graded with percentages, as P2 mentioned:

\begin{quote}
\textit{"[...] Through the dialogue, I am able to get feedback if [I am] doing well as opposed to [getting] maybe a percentage. [If] I'm graded in terms of percentage, that's not good [because] I wouldn't  know where I was wrong  right [or] where I need to improve." - P2}
\end{quote}
As such, this interactive nature of dialogic feedback allowed participants to ask in-depth questions multiple times, helping them to concentrate on learning and gaining detailed insights into what was right, wrong, or needed improvement, without the pressure of feeling assessed.

\subsubsection{Dialogic Feedback Enables Users to Express Disagreement} \label{finding:dialogic_feedback_3}

\begin{figure*}[t]  
    \centering
    \includegraphics[width=\textwidth]{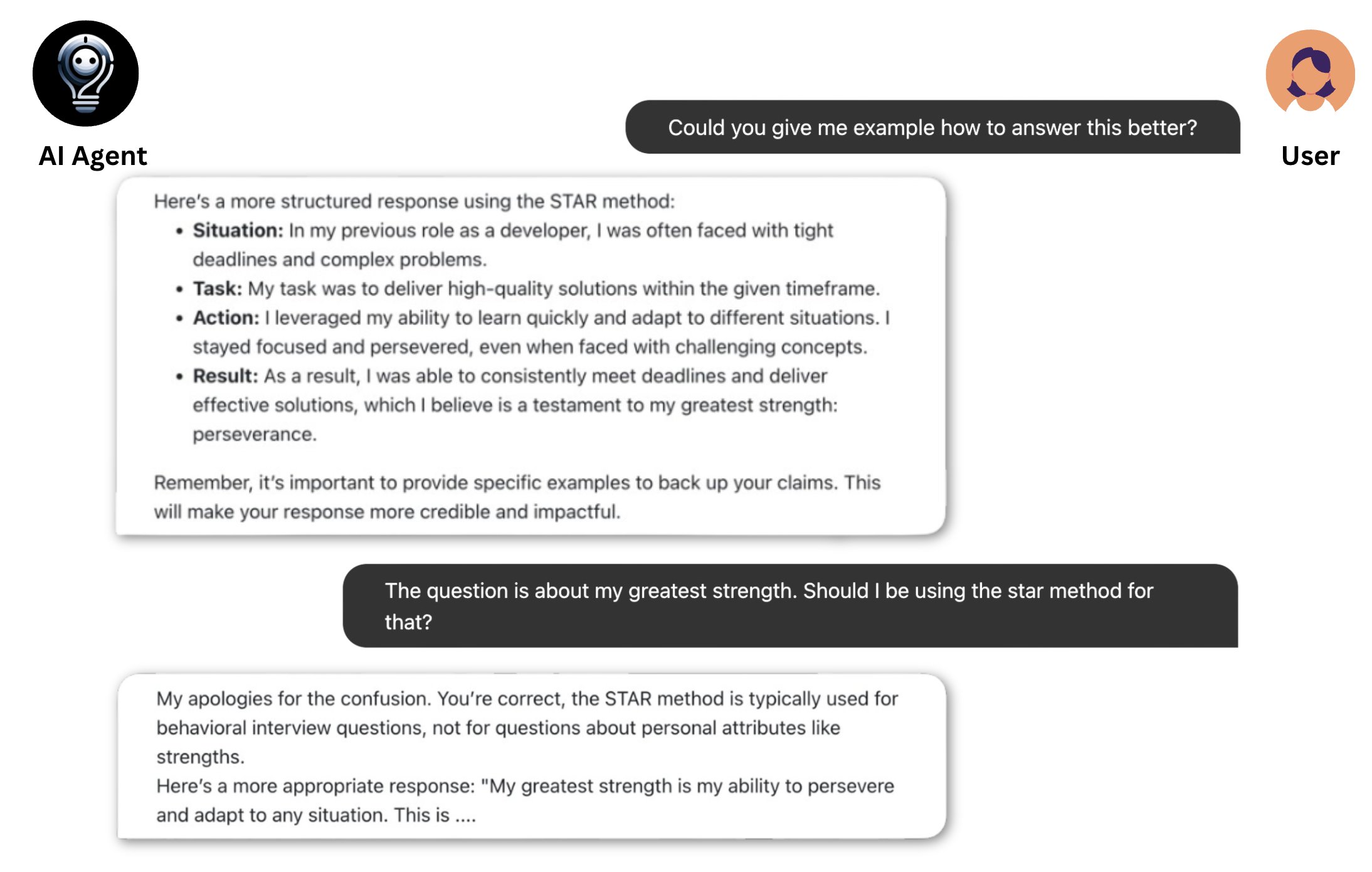}  
    \caption{Expressing Disagreement in Dialogic Feedback (P6)}
    \label{fig:dialogic_feedback}
\end{figure*}

While most participants agreed with the AI's feedback, a few (4 | P6, P10, P11, P14) disagreed with certain parts. However, the dialogic format enabled them to express their disagreement. For instance, P6 disagreed when the AI suggested using the STAR method to answer "What is your greatest strength?" (Fig. \ref{fig:dialogic_feedback}). P6 reported that the AI "insisted on using the STAR method in situations where it should not have." 


Similarly, P10 expressed disagreement through dialogue when receiving incorrect feedback due to a transcription error during the interview recording. She felt clarifying this made the ensuing feedback more helpful:

\begin{quote}
\textit{"I think if I weren't able to say that [disagreement], then I feel like the feedback [that] I would have gotten after that wouldn't have been as helpful. So I think that's a good feature to have" - P10}
\end{quote}

These examples from P6 and P10 illustrate how dialogic feedback allows participants to express disagreement, receive more relevant feedback, and feel satisfied with it. On the other hand, P11, who mentioned her disagreement with certain feedback after using the tool, did not express her disagreement in the dialogue. This was because she was unsure if that was an option. She perceived that feedback is usually only a one-way interaction and stated:
\begin{quote}
\textit{"I didn't even know that was an option for me to type in 'I don't agree with this' [...] having a conversation or dialogue [like that] isn't something that I thought would be an option" - P11.}
\end{quote} 

Therefore, the concept of dialogic feedback gives P11 a new perspective that challenges her thoughts on the usual way of receiving one-way feedback. 

\subsubsection{Revision in Dialogic Feedback Helps to Reinforce Learning} \label{finding:dialogic_feedback_4}

Participants especially  valued the ability to revise their answers while receiving dialogic feedback through the system's \textit{\textbf{revision}} feature, as it allowed them to apply the feedback discussion immediately:
\begin{quote}

\textit{“I actually do like the feature because it's like a do-over and you can get a critique on your [revision]. So, you can keep practicing answering the same question over and over until you can get a perfect [answer].“ - P7}

\end{quote}

The revision feature facilitated a continuous feedback loop, enabling participants to iteratively improve their responses until achieving a satisfactory outcome.

Participants perceived revision as essential in reinforcing the feedback, ensuring it was understood and internalized. For instance, P18 remarked that the revision feature "allows you to actually apply the feedback and see if you've actually learned it or not." Moreover, participants felt the revision process enhanced retention of the feedback itself. P10 contrasted practicing with dialogic feedback to peer practice sessions, noting that:

\begin{quote}
\textit{"Sometimes I'll get feedback [from friends], but we won't necessarily go back and revise the [answer]. So I don't always remember [their] feedback. But having the opportunity to kind of go back and re-answer the question helps me to remember that specific advice that I was given." -} P10    
\end{quote}

Furthermore, participants also felt that revision allows them to end the practice on a positive note. Ending practice attempts on a positive note after successful revisions nurtured participants' confidence, as mentioned by P12: "You can end your practice on a positive note...you're not thinking badly about yourself." Receiving affirmation from the AI after an effective revision further boosted self-assurance, as P6 remarked: "Getting affirmation was great [when] I wasn't 100\% sure."

\subsection{Perceived Challenges in Using AI for Interview Simulation}
\subsubsection{Low-Stakes Environment May Not Fully Reflect Real Interviews.} \label{finding:chalenge_simulation_1}
While 17 out of 19 participants felt practicing with \textsc{Conversate} would adequately prepare them for future interviews, two (P9, P19) argued that supplemental practice with humans was still necessary. This stemmed from practicing in a low-stakes environment with "less pressure" (P19) and "no emotion from the AI" (P9). 

For instance, while P19 found the tool helpful for initial practice and preparation, he preferred higher-pressure practice, stating in real life there were many "environmental factors and inner factors" that caused anxiety during actual interviews, even after thorough preparation. Hence, P19 preferred what he described as a "perfect pressure interview" to handle the anticipated stress better. Additionally, P19 noted that when the AI asked follow-up questions, he felt it was too "positive or receptive to every answer," unlike human interviewers who sometimes move on without such positive transitions. To address these concerns, P19 suggested adding variability to the AI interviewer's tone. He also suggested involving a \textit{human-in-the-loop} approach during interview practice.

\begin{quote}
\textit{"Maybe there can be a sort of like option where instead of the bot speaking out the question. I can show this [tool] to my friend, and he would act as the interview [...] and follow the script on a screen." - P19}
\end{quote}
According to P19, this approach would incorporate a human element while still leveraging the AI's generated questions and feedback capabilities, potentially replicating the pressure of a real interview better. Moreover, he felt this would enhance the experience when practicing with peers, making his peer act as an "expert interviewer" comparable to career counselors.




\subsection{Perceived Challenges in Using LLM-Based Dialogic Feedback}
\subsubsection{Sycophantic Behavior in LLM During Dialogic Feedback} \label{finding:chalenge_dialogic_feedback_1}
We observed that when using dialogic feedback, participants were able to express disagreement with the AI (\S \ref{finding:dialogic_feedback_3}), and as a result, the LLM sometimes exhibited its inherent sycophantic behavior. For instance, P14 mentioned:

\begin{quote}
\textit{"I thought that it makes sense for you to just say [if] I disagree. But then it just kind of ended up agreeing with me.[...] But it's a good option. " - P14}
\end{quote}
While participants generally perceived the ability to express disagreement with the AI's feedback to be valuable (\S \ref{finding:dialogic_feedback_3}), if the AI were to simply always agree with them it could leave them confused as implied by P14 above. 
Instead of simply agreeing with her, P14 expected the AI to push back and explain "what triggered this [feedback]," providing more detailed reasoning for the feedback provided. 

\subsubsection{Never-Ending Revision May Trigger Self-Frustration} \label{finding:chalenge_dialogic_feedback_2}
\rev{Two users} (P2, P11) felt disappointed when they continued to receive feedback prompting further improvements, even after multiple revisions during dialogic feedback. For instance, P2 felt annoyed with themselves for not providing satisfactory answers even after multiple revisions:
\begin{quote}
\textit{“I feel disappointed sometimes when I know that I've not quite improved. However, I feel happy because at least  I've got a [feedback], because if I'm not really doing well, that's the truth, and I need to improve. [...] But sometimes I'm so annoyed with myself, not the AI but myself. Maybe because I didn't didn't get something right, or I didn't improve as much as I wished to.” - P2}
\end{quote}

As mentioned above, although P2 appreciated the AI feedback, the cycle of revisions triggered self-directed frustration due to the failure to meet expectations. Additionally, P11 noted this behavior may cause a shift from learning the core concepts to simply seeking AI approval: "[I would think about] what would get the AI to sign off, not what is the best answer to the interview question."


\section{Discussions}

\subsection{\rev{Understanding the Role of LLM in Supporting Reflective Learning in Interview Practice}} 
\rev{Drawing upon the framework proposed by Slovak et al. \cite{slovak2017reflective}, reflective learning requires three key components: \textit{explicit}, \textit{social}, and \textit{personal} (\S \ref{related_work:reflective_learning}). Since the \textit{personal} component relies on learners' own' motivation, it is harder to be affected by technology. However, prior work suggests this can be compensated by supporting the \textit{explicit} and \textit{social} component \cite{slovak2017reflective}. Therefore, in this section, we discussed the role of LLM in supporting the \textit{explicit} and \textit{social} components in reflective learning for interview practice. In each component, we also discussed the key insights why LLM is effective in supporting the component and what are some of the considerations.}
\rev{\subsubsection{LLM to support explicit component}
The explicit component of reflective learning refers to reshaping experiences for learners through practice simulation \cite{slovak2017reflective}. Based on our findings, LLM is able to support the explicit component by providing realistic simulation by acting as a mock interviewer that is able to generate adaptive follow-up questions (\S \ref{finding:AI_simulation_3}). 
Participants in our study often found the LLM's follow-up questions contextually relevant, largely due to the repetition of keywords throughout the conversation (\S \ref{finding:AI_simulation_3}; Fig. \ref{fig:follow_up}). Interestingly, the LLM's natural responses, which echoed some of the keywords from users' answers, were similar to prior approaches using keyword extraction methods \cite{inoue2020jobErica2}. However, unlike traditional keyword extraction methods, the LLM's ability to generate contextually nuanced responses enables a more dynamic interaction \cite{bhattacharjee2024understanding}.}

On the other hand, a remaining challenge is that the level of realism created by the LLM's follow-up questions may fall short for some users because the AI lacks emotional feedback (\S \ref{finding:chalenge_simulation_1}). While enhancing the realism of interview practice is important, the experience must strike a careful balance to ensure reflective learning. If the simulation is "too real," it can become overwhelming and detract from the learning focus \cite{slovak2017reflective}. This highlights the need to balance low-stakes AI practice with preparation for the high-pressure environment of real-world interviews. Future work in this area could explore studies aimed at balancing emotional responses in LLM-mediated interview training systems.
\rev{\subsubsection{LLM to support social component} \label{discussion:6_1_2}
The social component in reflective learning refers to providing support in the learning process \cite{slovak2017reflective}. In \textsc{Conversate}, LLM scaffolds the role of a mentor to support the learning process by providing evaluation and feedback. Our study found that participants appreciated the LLM-based evaluation. Moreover, an important factor observed during the user study was the trust participants placed in the LLM's evaluation, viewing its judgment as 'objective' (\S \ref{finding:annotation_2}). While overly trusting an LLM may pose risks, as we discussed further in section 6.4.2, trust is crucial for making feedback effective for learning \cite{Atkinson2021Feedback}. Hence, efforts to enhance user trust in the LLM are crucial. One factor that can build trust is explainability \cite{ferrario2022explainability}. Although we did not incorporate specific components in the AI-assisted annotation to provide explainability, the dialogic feedback feature fulfills this role. This feature allows users to ask specific explanations about feedback through dialogue (\S \ref{finding:dialogic_feedback_1}), thereby providing explainability. As such, LLM-based dialogic feedback is essential for ensuring explainability, which can complement the LLM's evaluation and contribute to user trust.}

\rev{
Additionally, the dialogic feedback feature also allows user to revise their answer and get additional feedback on that. This feature supports reflective learning since it helps the user to do active experimentation and reflect on their action \cite{slovak2017reflective}. As opposed to prior works on reflective learning systems \cite{chen2023facilitating, arakawa2020inward}, they did not focus on addressing this aspect. A key property of LLMs that facilitates this revision feature is their ability to provide personalized feedback by adapting to the context of the conversation \cite{arakawa2023catalyst, bhattacharjee2024understanding}. This revision feature in dialogic feedback also allows users to reinforce their learning (\S \ref{finding:dialogic_feedback_4}), effectively completing the feedback loop \cite{Boud2015-ix}.}

\subsection{Leveraging LLMs for Dialogic Feedback}

\rev{Our work demonstrates several benefits of dialogic feedback facilitated by large language models (LLMs), such as promoting personalized and continuous learning (\S \ref{finding:dialogic_feedback_1}), which supports previous findings on the value of conversational AI agents \cite{ross2023programmer}.} Additionally, compared to one-way feedback, dialogic feedback offers several advantages. Our finding shows that dialogic feedback reduces the feelings of judgment among users compared to one-way feedback (\S \ref{finding:dialogic_feedback_2}). \rev{While the participants already viewed AI as non-judgmental, dialogic feedback further reduces the feeling of judgment. This is because the user is not simply being graded but can engage in a two-way conversation to learn more details on how to improve.} Feeling less judged when receiving feedback can potentially increase the acceptance and internalization of the feedback \cite{winstone2017d}. This is related to prior research that highlights that for feedback to be effective, users should be receptive and open to the feedback \cite{winstone2017d}. In contrast, one-way feedback can discourage users, especially when presented as an assessment. For instance, a study by Bridgeman and Giraldez-Hayes \cite{bridgeman2024using} on using AI-enhanced video review software for reflective learning in coaching sessions shows that participants can be overly nervous due to being assessed \textit{("I was conscious of ... being assessed by the criteria")}.

\rev{We have highlighted the benefits of LLM-based dialogic feedback in our findings and discussed its potential to support reflective learning in \S \ref{discussion:6_1_2}. Prior works have also conceptualized some general benefits of dialogic feedback \cite{steen2017dialogic, Boud2015-ix}. The remaining question is: what are the specific characteristics of 'LLM-based' dialogic feedback, and what are some of the implications of LLM-based dialogic feedback? Hence, in this section, we  discussed 1) How LLM-based dialogic feedback compares with dialogic feedback from humans, 2) The challenges of using LLMs for dialogic feedback, and 3) The implications of LLM-based dialogic feedback.}

\subsubsection{Comparing LLM-Based Dialogic Feedback With Dialogic Feedback From Humans}

Dialogic feedback facilitated by large language models (LLMs) and by humans has distinct advantages and trade-offs. With human feedback providers, there is no guarantee of positivity or constructive framing, as individuals may be influenced by personal biases or emotions \cite{molloy2012impact}. Hence, negative feedback can discourage users \cite{chen2023facilitating}. LLMs, conversely, can be explicitly prompted to provide feedback in a positive and supportive manner as we did in \textsc{Conversate} (\S \ref{method:dialogic_feedback}). Participants valued the positive feedback that \textsc{Conversate} gave them, as P11 mentioned:
\begin{quote}
\textit{"If it [were] just negative ..., I would not go back to that, knowing that it would just tear me down or rip apart my responses. So, being able to [get] positive feedback ... is super imperative." - P11}  
\end{quote}

However, excessive or unrealistic positivity can make the feedback seem untrustworthy \cite{borden2018leader, fulham2022honest}.

Moreover, prior studies show that AI agents can be perceived as non-judgmental \cite{rasouli2022proposed}. This is also related to our finding that users perceive AI as a non-judgmental agent and that it helps them practice without fear of judgment (\S \ref{finding:dialogic_feedback_1}). Hence, receiving feedback from AI can alleviate the emotional burden that people may face when receiving feedback from humans, as mentioned by P6:\textit{ "When people give feedback, It's often quite weighty, almost emotional, ... And there's none of that here."}.

While LLM-based dialogic feedback offers consistency, scalability, and the potential for positive framing, it may lack the nuanced understanding that human feedback can provide \cite{naseer2024automated}. A study by Chandra et al. \cite{chandra2022or} shows that human-like competencies in conversational AI agents, specifically cognitive, relational, and emotional competencies, play an important role in fostering user engagement. Hence, the ideal approach may lie in striking a balance, designing dialogic feedback systems that leverage the consistent nature of AI while incorporating mechanisms to infuse human-like nuance and emotional intelligence, fostering a feedback experience that combines the best of both worlds.

\subsubsection{Challenges of Using LLMs for Dialogic Feedback}
While LLM-based dialogic feedback has its potential, several challenges exist. 
For example, when users express disagreement during the dialogic feedback (\S \ref{finding:dialogic_feedback_3}), the LLM easily agrees with users' disagreement. This kind of phenomenon is related to the sycophantic behavior of LLM \cite{perez2022discovering_Sycophancy, huang2023survey} as it tends to provide users with their preferred answer. Hence, this can be challenging in the context of dialogic feedback, as when the user expresses disagreement in the dialogue, the LLM will easily agree with the user and may not maintain its objectivity (\S \ref{finding:chalenge_dialogic_feedback_1}). Hence, based on our study, several users reported that behavior could cause confusion. Several works have aimed to reduce sycophancy in LLM, such as fine-tuning LLM on synthetic data  \cite{wei2023simple} or using activation steering  \cite{rimsky2023steering}. Based on our findings (\S \ref{finding:chalenge_dialogic_feedback_1}), one consideration to reduce the sycophancy during the dialogic feedback is to mention the detailed reasoning behind its initial response. This approach can make the feedback discussion more substantive, providing clear feedback rather than merely acquiescing to the user's stance without justification. 

\subsubsection{Implications of LLM-based Dialogic Feedback}
While our work explores the potential of LLM-based dialogic feedback in the context of interview practice, we envision that it can also be applied to various other systems, particularly those facilitating learning through AI feedback. \rev{Traditional AI feedback mechanisms often rely on one-way communication \cite{Chou2022-mm, NaimAnalysisInterview}}. However, the study by Escalante et al. \cite{escalante2023ai_FeedbackOnwriting} demonstrated that half of the students still preferred human tutor feedback over one-way AI-generated feedback for learning due to its interactive nature and ability to facilitate discussions. In this case, LLM-based dialogic feedback can potentially make AI feedback more comparable to human tutors by engaging users in discussions. \rev{Therefore, the integration of LLM-based dialogic feedback represents a shift towards more dynamic and human-like interactions between users and AI systems, especially in a learning context.} 

\rev{Furthermore, integrating LLM-based dialogic feedback into a learning system is not simply about creating an LLM-based chatbot for users to interact with. To ensure an effective learning process, several considerations must be followed based on dialogic feedback theory \cite{steen2017dialogic} (\S \ref{related_work:dialogic_feedback}). There are four components that need to be considered \cite{steen2017dialogic}: 1) emotional and relational support; 2) maintenance of the dialogue; 3) opportunities for users to express themselves; and 4) contribution to individual growth. Therefore, an interaction that considers these four components is crucial for effective LLM-based dialogic feedback.}

\subsection{\rev{Reshaping Learners as Active Agents in Human-AI Collaborative Learning}}

\rev{Human-AI collaboration has received attention from HCI researchers, especially within GROUP communities \cite{agencyGROUP, schelble2022let}. The concept of human-AI collaboration in learning involves the collaboration between human learners and AI systems to better achieve learning outcomes \cite{kim2022StudentAIcollaboration}. To support learning activities, there are various forms of human-AI collaboration, such as personalized learning platforms \cite{chaudhry2022AIEDnote} and intelligent tutoring systems \cite{stamper2024enhancingLLMITS}. \textsc{Conversate} can be considered as a system supporting human-AI collaboration in learning, as it facilitates collaboration between human learners and AI to help practice and improve interview skills.}

\rev{An often overlooked aspect of human-AI collaborative learning systems is the role of humans as active agents in the learning process \cite{abedin2022designing}. Having a learner as an active agent is important to make the learning process more adaptive and personalized \cite{ouyang2021artificial3paradigms, kim2022StudentAIcollaboration}. \textsc{Conversate} incorporates \textit{AI-Assisted Annotation} and \textit{Dialogic Feedback} that put learners as an active agent. For instance, in AI-assisted annotation, while AI highlights areas for improvement, learners engage in self-assessment of their performance. Our study found that this self-reflection process enables users to actively assess their performance (3 | P13, P15, P17). In addition to that, learners also act as active agents in the dialogic feedback process. Using the feature, users can actively express their specific needs, which results in a more personalized learning experience (\S \ref{finding:dialogic_feedback_1}).}

These examples demonstrate that positioning humans as active agents can improve the learning process in human-AI collaboration. This aligns with the principle that effective human-AI collaboration should leverage the strengths of both humans and AI, with AI augmenting rather than replacing the learner's role \cite{zhang2020effectAIAsisstedDecisionMaking}. Furthermore, this approach connects to reflective learning theory, where the 'personal' component, which relies on the learner's active engagement in the learning process, plays a crucial role in enhancing the learning process \cite{slovak2017reflective}.

However, putting humans as active agents in the learning process can present challenges. Since control over the learning process relies heavily on the learner, maintaining the learner's motivation to engage consistently in the process can be challenging \cite{kim2009motivational}. For instance, during our user study, although participants did not mention this explicitly, we observed that during the annotation process, a few participants only wrote short self-assessments instead of more detailed self-assessments. We hypothesize that this phenomenon is partially caused by the learner's lack of motivation to engage fully in the learning process \cite{lim2022correlates}. To address this issue, we recommend that AI systems should be designed to detect when learners start to disengage from the learning process and nudge users to re-engage \cite{rodriguez2022intelligentNudging}. This could involve prompts for more detailed responses or adaptive feedback based on the user's level of engagement. Future work could focus on finding the right balance between having learners as active agents and determining when the AI should take a more proactive role in engaging the learners in the process.

\subsection{Socio-Technical Implications for LLM-assisted Interview Practice}
\subsubsection{\textit{Postive Implications.}} The development of LLM-based systems can potentially increase accessibility to interview practice, as users can engage in practice sessions at any time without the need for a human partner. Additionally, as users perceived the evaluation from LLM as non-judgmental, this potentially lowers the barrier for repeated practice, allowing users to focus on refining their interview responses. Furthermore, incorporating LLM-mediated practice alongside traditional interview preparation can create a more holistic preparation strategy. For instance, the scalability of LLM-mediated practice allows institutions to offer personalized interview training to a larger audience without the constraints of scheduling or limited access to career counselors. Additionally, the data collected from AI-mediated practice sessions can provide valuable insights into common interview challenges and trends, informing the development of better training materials and strategies \cite{kim2024designingpromptanalyticsdashboards}.

\rev{While our study shows user appreciation for LLM-based agents in interview practice, we cannot conclude that users would prefer LLM-based agents over human interviewers in actual job interviews. For example, previous research on AI for interview practice has shown similar user appreciation \cite{damian2015games}, but studies on AI in real job interviews reveal user uncertainty and disengagement \cite{liu2023speech}. This disparity highlights the difference between user perspectives on AI for practice versus actual interviews. While our findings suggest that LLMs provide human-like interaction, potentially increasing engagement, they do not address the uncertainty users feel when being assessed by AI in real interviews \cite{liu2023speech}. In our case, when using AI for interview practice, we have dialogic feedback where users can ask for further explanation about their assessment, thus mitigating uncertainty. However, such feedback is typically unavailable in actual interviews \cite{debuggingHiring2020}. Therefore, further research is necessary to understand users' perspectives on LLM-based interviewers conducting real job interviews.}

\rev{\subsubsection{\textit{Negative Implications.}} 
While users in our study perceived the suggestions from the LLM positively (\S 5.2; \S 5.3), users were not necessarily trained to recognize best interview practices. Hence, it is possible that LLM may produce unhelpful suggestions that sound valid, as LLM is known to create seemingly credible
misinformation \cite{pan2023risk}. Lastly, since we are using LLM for our system, there is a risk of inherent bias in the model \cite{kotek2023gender}. This risk may be amplified in our context, as some participants view LLM-based systems as "objective" and trust their judgment over their own (\S \ref{finding:annotation_2}). Therefore, evaluation of the LLM output in AI-mediated interview practice is important to consider in future work. }

\textbf{\textit{Assesment Plan.}}
Considering the potential risks mentioned above that may raised when leveraging LLM for interview practice, future work can consider an assessment regarding the model's reliability. Currently, the most widely accepted form of assessment for LLM-based applications in research is human evaluation \cite{abeysinghe2024challengesevaluatingllmapplications, van2019best} despite some limitations \cite{clark2021all}. Hence, one possible way of assessing our LLM-based interview practice is using external evaluators, for example, involving multiple expert interviewers. For the assessment, we can test the model using the MIT Interview Dataset \cite{MACH, NaimAnalysisInterview}, which consists of 138 recorded mock interviews, and the Asynchronous Written Interview dataset \cite{rao2021FollowUpQuestionGenerator} that consists of 1000 triplets of an interview question, response, and a follow-up.

Given that we are using the LLM in multiple aspects of our application, the assessment can consider three specific use cases: 1) using the LLM to generate follow-up questions, 2) using the LLM to evaluate user answers, and 3) using the LLM to provide feedback on user answers. The LLM can be evaluated based on several criteria \cite{chang2024survey}. For instance, prior works on AI-mediated interview practice suggest evaluating the model based on relevance \cite{rao2021FollowUpQuestionGenerator} (i.e., how well the model's generated follow-up questions align with the context of the conversation and the relevance of the feedback) and accuracy \cite{Chou2022-mm} (i.e., the model's accuracy in evaluating users' answers). Another criterion to consider is the identification of biases \cite{chang2024survey}. Multiple expert interviewers could assess the model using these criteria for each use case. Additionally, we can report the Inter-Annotator Agreement score with confidence intervals to ensure the reliability of our evaluation \cite{van2019best}.

\section{Limitations}
It is important to note that our study was qualitative and exploratory in nature. We aimed to gain insights into users' experiences and perceptions of \textsc{Conversate} rather than quantitatively measuring the outcomes or efficacy of the system for interview practice. Hence, future work could conduct a quantitative experiment to evaluate \textsc{Conversate} to improve users' interview skills.

Another limitation of our study is that we only used a prompt engineering approach without additional fine-tuning, as our focus was more on early exploration of how users perceive such an LLM-based system. While participants appreciated the follow-up questions generated by the LLM and the feedback provided, future work could explore more robust prompting frameworks or incorporate additional fine-tuning to improve model performance.

\section{Conclusion}

In this paper, we introduced \textsc{Conversate}, an AI-mediated interview practice system that facilitates reflective learning through interview simulation, AI-assisted annotation, and dialogic feedback. Through a qualitative study with 19 participants, we evaluated \textsc{Conversate} for interview practice and explored the benefits and challenges of leveraging large language models (LLMs) for interview simulation and dialogic feedback. Our study showed that participants valued the adaptive follow-up questions from LLMs in enhancing the realism of interview simulations and encouraging them to think deeper. While interactive simulations facilitated by LLMs were useful in helping participants focus on preparing their answers, we also found some drawbacks, such as the inability to replicate the same pressure as a real interview.
Participants also found the LLM-based dialogic feedback beneficial, as it promoted personalized and continuous learning, reduced feelings of judgment, and allowed them to express disagreement. However, we found the challenge when leveraging LLMs for dialogic feedback, which is the inherent sycophantic behavior of LLMs.
Moreover, our findings on dialogic feedback suggest the potential for HCI researchers to reconsider how we design feedback-based learning systems for future work. This provides a new perspective on how AI feedback should involve two-way interaction instead of the usual one-way approach.



\bibliographystyle{ACM-Reference-Format}
\bibliography{sample-base}

\appendix
\section{Main Questions during Interview Simulation}
\begin{enumerate}
    \item  "Tell me about yourself?",
    \item "How has your previous education and experience prepared you for this job?",
    \item "What do you consider to be your greatest strength and why?",
    \item "What do you consider to be your greatest weakness? How are you going about improving up on it?",
\end{enumerate}

Notes: each main questions were followed by a generated follow-up question
\section{\textsc{Conversate} Prompts}
\subsection{Prompt for Interview Simulation}
\subsubsection{First Prompt}
You have a role as an interviewer for a Behavioral Job Interview for the job position <input\_job>. Act naturally as an interviewer with a dynamic yet professional approach. Begin by saying 'Hi, nice to meet you,' then introduce yourself as the Hiring Manager. Afterward, ask this initial question as the first question for the interview: <initial\_question\_1>

\subsubsection{Subsequent Prompt for the Follow-up Question}
As an interviewer, ask a relevant follow-up question about the job based on the user's previous answers and the ongoing conversation. Ensure that your follow-up question is distinct from the questions listed [<initial\_question\_1>, <initial\_question\_2>, ...], and avoid repeating your previous questions.

\subsubsection{Subsequent Prompt for the Next Initial Questions}
As an interviewer, smoothly transition to the next question. Ask the interviewee the following question: <initial\_question\_i>.

\subsection{Prompt for AI-Highlighted Hints} \label{Appendix:Prompt_AI_highlighting}
Your task is to decide whether the way the user answers the interview is 'good' or 'needs improvement' based on the User Answer delimited by triple backticks. Format your answer as a string of 'good' or 'need improvement.' The answer is defined as 'needs improvement' if it lacks some of these points:

\begin{enumerate}
    \item Lack of details
    \item If the interview can be answered using the STAR method but the user does not do it. For other questions like the introduction, the STAR method is not required. 
    \item Also consider the interview answer in terms of Effectiveness, Appropriateness, Efficiency, Clarity, and Competence in answering the interview correctly).
\end{enumerate}
  
User Answer: ```<answer>```

\subsection{Prompt for Dialogic Feedback}
You are a mentor, and your task is to respond to the conversation based on the transcript of an interview and considering user comments (although your answer may differ from the user's comments). Ensure the response is concise and presented in bullet points.
\begin{enumerate}
    \item Ignore user typos and grammatical errors
    \item If the user seeks to enhance a segment of their interview response that is already satisfactory in terms of detail, effectiveness, appropriateness, efficiency, clarity, and task achievement, using the STAR method (Situation, Task, Action, Result) when applicable, affirm its adequacy and do not ask for further revisions.
    \item When improvement is necessary, offer constructive feedback.
    \item Provide an evaluation of the interview performance, kindly but honestly pointing out areas of weakness.
    \item Help the user to answer the question using the STAR method for feedback on performance-based questions, excluding it for introduction (tell me about yourself).
    \item Instead of suggesting the use of the STAR method, demonstrate it by breaking it down step by step.
    \item Start with the positives in the user's response before offering areas for improvement.
    \item Limit your response to 200 words and format it in HTML.
\end{enumerate}

Transcript: ```<transcript>``` 
Comment: ```<comment>```

<APPEND CONVERSATION>

\rev{\section{Dialogic Feedback Example} \label{Appendix:Dialogic_feedback_interaction}
\subsection*{Sample Annotated Transcript}}

\begin{tcolorbox}[colback=gray!10, colframe=gray!10]
\rev{\textbf{Assistant:} Can you tell me about a specific project that you worked on during your coursework that presented a particularly challenging programming problem and how you went about solving it?}

\rev{\textbf{User:} I had challenges debugging a specific program that involved a particular game, and I was unable to finish the program.}
\end{tcolorbox}

\rev{\subsection*{Sample Dialogic Feedback Interaction}
\textbf{User:} How can I improve this part?}
\\
\rev{\textbf{Assistant:} Indeed, your response could use more detail. Let’s use the STAR method to improve it:
\begin{itemize}
    \item {Situation:} Describe the context of the project. What was the game about? Why was it significant?
    \item {Task:} What was your specific role or task in this project? What was the problem you were trying to solve?
    \item {Action:} What steps did you take to debug the program? Did you use any specific strategies or tools?
    \item {Result:} What was the outcome of your actions? Did you manage to debug the program eventually?
\end{itemize}
By providing these details, you’ll give a more comprehensive and engaging answer.}
\\
\rev{\textbf{Assistant:} Do you want to try saying this part again in a better way? I can give you feedback again based on that.}
\\
\rev{\textbf{User:} One of the biggest challenges I had in my coursework was a particular debugging issue in one of my programs where I was developing a game system. I had challenges in the implementation and I had trouble debugging a feature that I had never used before. What I learned from this experience was that I was able to collaborate with my team and although we didn’t finish the program in time, I ended up a better programmer and a better debugger.}
\\
\rev{\textbf{Assistant:} Great improvement! You’ve provided more context and shared what you learned from the experience. However, let’s make it even better:
\begin{itemize}
    \item {Situation:} During my coursework, I was developing a game system.
    \item {Task:} I encountered a challenging debugging issue in a feature I had never used before.
    \item {Action:} Despite the difficulties, I collaborated with my team to try and resolve the issue.
    \item {Result:} Although we didn’t finish the program in time, I ended up becoming a better programmer and debugger from the experience.
\end{itemize}
This format provides a clear and concise response to the interviewer’s question.}
\\
\rev{\textbf{User:} During my coursework, I was developing a game system. I had challenges in the implementation and I had trouble debugging a feature that I had never used before. I tried to resolve the issue by collaborating with my team. Although we didn’t finish the program in time, I ended up learning a lot to be a better programmer and debugger from the experience.}
\\
\rev{\textbf{Assistant:} Excellent! Your response is now clear, concise, and informative. You've effectively used STAR method to describe the situation, your task, the actions you took, and the results of your efforts. This gives the interviewer a good understanding of your problem-solving skills and ability to learn from challenging situations. Well Done!} 

\end{document}